\documentclass[11pt]{article}
\pdfoutput=1
\usepackage{amsmath}
\usepackage{amssymb}
\usepackage{graphicx}% see geometry.pdf on how to lay out the page. There's lots.
\usepackage{xcolor}
\usepackage{customcommands}
\usepackage{jheppub}
\usepackage{subfigure}
\usepackage{bm}
\usepackage{ulem}%I added this package just for editing purposes

\title{Towards Bulk Metric Reconstruction from Extremal Area Variations}

\author[1,6]{Ning Bao,}
\author[2,3]{ChunJun Cao,}
\author[4]{Sebastian Fischetti,}
\author[5]{and Cynthia Keeler}

\affiliation[1]{Berkeley Center for Theoretical Physics, University of California Berkeley, Berkeley, CA, 94720, USA}
\affiliation[2]{Walter Burke Institute for Theoretical Physics, California Institute of Technology, Pasadena, CA, 91125, USA}
\affiliation[3]{Joint Center for Quantum Information and Computer Science, University of Maryland, College Park, MD, 20742, USA}
\affiliation[4]{Department of Physics, McGill University, Montr\'eal, QC, H3A 2T8, Canada}
\affiliation[5]{Department of Physics, Arizona State University, 
Tempe, AZ, 85287, USA}
\affiliation[6]{Computational Science Initiative, Brookhaven National Laboratory, Upton, New York, 11973}

\emailAdd{ningbao75@gmail.com}
\emailAdd{ccj991@gmail.com}
\emailAdd{fischetti@physics.mcgill.ca}
\emailAdd{keelerc@asu.edu}

\abstract{
The Ryu-Takayanagi and Hubeny-Rangamani-Takayanagi formulae suggest that bulk geometry emerges from the entanglement structure of the boundary theory.
Using these formulae, we build on a result of Alexakis, Balehowsky, and Nachman to show that in four bulk dimensions, 
the entanglement entropies of boundary regions 
of disk topology 
uniquely fix the bulk metric in any region foliated by the corresponding HRT surfaces.  More generally, for a bulk of any dimension~$d \geq 4$, knowledge of the (variations of the) areas of two-dimensional boundary-anchored extremal surfaces of disk topology uniquely fixes the bulk metric wherever these surfaces reach.  This result is covariant and not reliant on any symmetry assumptions; its applicability thus includes regions of strong dynamical gravity such as the early-time interior of black holes formed from collapse.  While we only show uniqueness of the metric, the approach we present provides a clear path towards an \textit{explicit} spacetime metric reconstruction.
}
\begin{document}

\maketitle

\section{Introduction}
\label{sec:intro}

The AdS/CFT correspondence provides an indirect definition of a nonperturbative, background-independent theory of quantum gravity: the fundamental microscopic degrees of freedom of the boundary field theory \textit{are} the microscopic degrees of freedom of the bulk theory of quantum gravity.  Fruitfully exploiting this observation requires understanding how these degrees of freedom, which are well-understood from the perspective of nongravitational quantum field theory, reorganize themselves (in an appropriate limit) into a manifestly local gravitational theory.  This challenge is broadly termed ``bulk reconstruction''.

The Ryu-Takayanagi (RT) and Hubeny-Rangamani-Takayanagi (HRT) proposals for holographic entanglement entropy~\cite{RyuTak06,HubRan07,LewMal13,DonLew16} provided a key insight into this reorganization. These proposals state that the entanglement entropy of a subregion~$R$ of the CFT is given by
\be
\label{eq:HRT}
S[R] = \frac{\Area[X[R]]}{4G_N \hbar},
\ee
where~$X[R]$ is the minimal-area codimension-two extremal%
\footnote{As is conventional in the AdS/CFT community, we use the misnomer ``extremal'' to refer to surfaces that are merely stationary points of the area functional, even if they do not extremize it.} %
surface homologous to~$R$ (and we are omitting~$1/N$ corrections).  Since the HRT formula relates boundary entanglement and bulk geometry,~\cite{Van09,Van10} suggested that the entanglement structure of the boundary state must play a crucial role in the emergence of the bulk.  This observation led to great strides in perturbative bulk reconstruction: thanks to additional insights from quantum error correction~\cite{AlmDon14,DonHar16}, we now understand how (approximately) local bulk operators \textit{on a fixed background geometry} are encoded in the operator algebra of the boundary field theory (see e.g.~\cite{Har18} for some review).  This fixed background geometry identifies a subspace of the CFT Hilbert space consisting of all those states with that same dual geometry (to leading order in~$1/N$).  Borrowing from the language of quantum error correction,~\cite{AlmDon14} refers to this subspace as a ``code subspace''. In this language, the bulk geometry is the code subspace and the reconstruction of~\cite{DonHar16,FauLew17} (as well as the earlier~\cite{HamKab06,HamKab05,Kab11}) is the reconstruction of local operators on a code subspace.

However, a full understanding of bulk reconstruction must be nonperturbative: that is, it requires not just reconstruction of operators on a given code subspace, but also an understanding of which CFT states even belong to a code subspace and how to determine what that code subspace is.  In other words: which CFT states correspond to a dual bulk geometry, and how does this geometry emerge from boundary degrees of freedom?  Here we endeavor to answer this second question: how, precisely, does the bulk spacetime arise from the boundary?

Several partial results in this direction exist, which we now highlight along with their limitations.  Motivated by the expectation that the bulk should emerge from the entanglement structure of the boundary state,~\cite{LasMcD13,FauGui13,SwiVan14,Czech:2016tqr,Mosk:2016elb,Faulkner:2017tkh} showed that for perturbations of the vacuum, any dual geometry which satisfies the HRT formula must satisfy the (perturbative) Einstein equation.  Since the HRT formula~\eqref{eq:HRT} assumes the Einstein equation in the first place, these results can be interpreted as a nontrivial consistency check on it; however, their purpose is to use~\eqref{eq:HRT} to constrain the bulk \textit{dynamics}, rather than to give a kinematic reconstruction of the bulk geometry.  Such a kinematic construction for static slices of~$(2+1)$-dimensional bulk geometries was achieved in~\cite{Czech:2014ppa} using ideas from differential entropy and hole-ography~\cite{Balasubramanian:2013rqa, Balasubramanian:2013lsa,Myers:2014jia,Czech:2014wka}; however, the no-go theorem of~\cite{Engelhardt:2015dta} shows that any covariant generalization of this construction cannot rebuild the metric everywhere extremal surfaces reach.  In particular, it cannot rebuild the metric in regions of strong gravity, which are perhaps the most interesting bulk regions to probe.  More generally but less explicitly,~\cite{KabLif18} sketches a formal way of identifying approximately local bulk operators by studying the modular Hamiltonians of all possible boundary causal diamonds, and then using these to fix the bulk metric. In a different direction, there have been attempts to ``discretize'' a slice of the bulk geometry into a tensor network that reproduces the entanglement structure of the boundary state~\cite{Swingle:2009bg,Swingle:2012wq,PasYos15,BaoPen18,MilVid18,BaoPen19}, but it is unclear how to covariantize such a construction in order to interpret it as an approximation of a general, dynamical, bulk geometry.

In a more direct approach that doesn't rely on entanglement, one could assume the bulk gravitational dynamics and na\"ively ``integrate in'' the equations of motion from the boundary.  However, hyperbolic boundary-value problems of this type are generically ill-posed. Even when they do yield a solution, it is too coarse-grained: the boundary conditions supplied to the classical bulk equations of motion are just one-point functions on the boundary, while the bulk should be sensitive to the microscopic boundary state.  This excessive coarse-graining is manifest in the fact that ``integrating in'' from a boundary region~$R$ in this way can recover at most the causal wedge of~$R$, which generically is a proper subset of the entanglement wedge to which~$R$ is expected to be dual.  A more refined construction is that of light-cone cuts developed in~\cite{EngHor16,EngHor16b}, which determines the bulk \textit{conformal} metric from singularities in~$n$-point functions in the dual field theory.  This approach has the advantage that it serves as a diagnostic on when a CFT state admits a dual geometry~\cite{EngFis17,EngFis17b}, but it too cannot recover any of the bulk geometry outside of the causal wedge.  Relatedly but much more restrictively, approaches like those of~\cite{Hammersley:2006cp,Hammersley2008,Bilson:2008ab,Bilson:2010ff} assume a bulk with a high degree of symmetry and then try to recover the remaining degree(s) of freedom; more generally,~\cite{RoySar18} constructs the conformal metric on static time slices from two-point functions of heavy operators.

We therefore conclude that a background-independent prescription for explicitly reconstructing a general bulk metric (when one exists) from boundary objects is still lacking.  We expect these boundary objects to be entanglement entropies (or something closely related to them), so the HRT formula~\eqref{eq:HRT} naturally leads to a purely geometric question: do the areas of extremal surfaces anchored to the boundary of a manifold uniquely determine its geometry, and if so, how is this geometry recovered from these areas?  This question is in the class of so-called boundary rigidity problems (see~\cite{Porrati:2003na} for an early review in the context of AdS/CFT), and notable uniqueness results include the rigidity of simple two-dimensional Riemannian manifolds from the lengths of boundary-anchored geodesics~\cite{2003PestovUhlmann,CrokeReview,croke1991,Croke2005,Stefanov2008} as well as the rigidity of three-dimensional Riemannian manifolds from the areas of minimal boundary-anchored two-dimensional surfaces~\cite{AleBal17} (under nontrivial additional assumptions on the geometry).  Constructive results are more difficult, but in the case of two-dimensional manifolds they include~\cite{Monard2014,Monard2015,Krishnan2010OnTI,UhlmannPestov2004}, while perturbative results have been applied in e.g.~seismology and medical imaging~\cite{DahlenTromp,Gullberg2000,Malecki2014}.

The main result of this paper, therefore, is to extend the uniqueness result of~\cite{AleBal17} to higher dimensions (and Lorentzian geometries)\footnote{We should note that while~\cite{AleBal17} presents a mathematically rigorous theorem, our argument relies on some physically reasonable but not rigorously proven assumptions in Section~\ref{subsec:isothermal}; for this reason, we refer to our result as an argument rather than a rigorous proof.}: specifically, we will show, under an appropriate set of assumptions given explicitly below, that second area variations of two-dimensional boundary-anchored extremal surfaces are sufficient to uniquely fix the metric (up to diffeomorphism) everywhere in a neighborhood of these surfaces.  The ambient geometry is assumed to have any dimension~$d \geq 4$ (and in fact can have any signature), which substantially simplifies a portion of the argument as compared to the~$d = 3$ case studied in~\cite{AleBal17}.  However, for a technical reason we always take the extremal surfaces to be two-dimensional, so it is clear that the~$d = 4$ case is of most immediate relevance to AdS/CFT (since for~$d = 4$ two-dimensional surfaces are also codimension-two).  We would like to emphasize, however, that our argument does not require the bulk metric to have any particular symmetry; it applies to dynamical geometries, including a portion of the region behind the early-time event and apparent horizons of a black hole formed from collapse.

This uniqueness result is not particularly surprising, as the problem of determining a bulk metric from the areas of \textit{all possible} boundary-anchored extremal surfaces is very overconstrained, and the holography community often implicitly assumes that such a result must be true.  From this perspective, our result simply puts this expectation on much firmer footing.  More interestingly, our result is \textit{almost} completely constructive: a key non-constructive part of our argument is the invocation of the uniqueness result~\cite{AlbGui13}, which applies to inverse boundary-value problems for elliptic differential operators; roughly speaking, such problems ask if an elliptic partial differential operator~$L$ be recovered from the boundary value and normal derivative of all solutions to the equation~$L \phi = 0$ on some compact domain.  Though to our knowledge the particular problem considered in~\cite{AlbGui13} does not have a constructive solution, many closely related problems do, so we suspect that a constructive version of~\cite{AlbGui13} should exist.  Moreover, our argument requires only knowledge of first and second variations of the areas of boundary-anchored extremal surfaces; converting to AdS/CFT parlance, this means we only require access to first and second variations of the entanglement entropies of boundary subregions.  Such variations are much simpler to control from a boundary perspective than the full entanglement entropies themselves, and are significantly more tractable than requiring access to the full modular Hamiltonian, as in some other holographic approaches to bulk metric reconstruction (e.g.~\cite{KabLif18}).

Because some details of the argument, which broadly follows the structure of~\cite{AleBal17}, are rather technical, in Section~\ref{sec:overview} we present an overview of it.  In that same Section we also list our assumptions and provide some discussion of their interpretation and strength.  Section~\ref{sec:details} contains the argument itself, presented in a way that fleshes out the outline provided in Section~\ref{sec:overview}.  Finally, Section~\ref{sec:conc} summarizes the result and lists open directions, including the ingredients necessary to make our argument fully constructive, higher-curvature and quantum corrections, higher dimensional surfaces, and generalizations to probe even deeper into the bulk.

%do we want to say anything about extremal surface that go behind bh horizon for certain dynamical solns here
%should we emphasize on the local reconstruction from partial data: this is a more powerful result

\subsubsection*{Preliminaries}

We will consider~$d$-dimensional Lorentzian spacetimes~$(M, g_{ab})$ with boundary~$\partial M$ on which the induced metric is~$h_{ab}$.  Two-dimensional spacelike surfaces embedded in $M$ will be called~$\Sigma$, and their induced metric will be written as~$\sigma_{ab}$.  The projector onto the normal bundle is defined to be~$P_{ab} \equiv g_{ab} - \sigma_{ab}$.  The extrinsic curvature of~$\Sigma$ is given by
\be
{K^a}_{bc} = -{\sigma_b}^d {\sigma_c}^e \grad_d {\sigma_e}^a.
\ee
Their mean curvature is the trace~$K^a \equiv \sigma^{bc} {K^a}_{bc}$, which vanishes if and only if~$\Sigma$ is extremal.

Lower-case letters from the beginning of the Latin alphabet (e.g.~$a$,~$b$,~$c$) will be used exclusively as abstract indices.  Lower-case letters from the middle of the Greek alphabet (e.g.~$\mu$,~$\nu$) will be used for spacetime indices, and range from~$1$ to~$d$; lower-case letters from the beginning of the Greek alphabet (e.g.~$\alpha$,~$\beta$) will label coordinates on~$\Sigma$, and range from~$1$ to~$2$; lower-case letters from the middle of the Latin alphabet (e.g.~$i$,~$j$) will label directions normal to~$\Sigma$, and range from~$3$ to~$d$.

\section{Overview}
\label{sec:overview}

The purpose of this Section is to give a broad overview of the structure of our argument.

\subsection{Assumptions}
\label{subsec:assump}

\begin{figure}[t]
\centering
\includegraphics[page=1,width=0.35\textwidth]{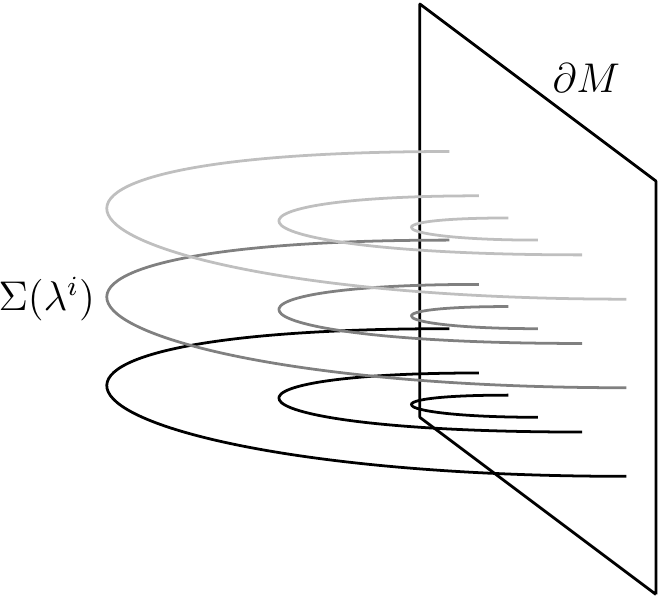}
\caption{An illustration of the foliation~$\Sigma(\lambda^i)$ of extremal surfaces we consider (for clarity we suppress a dimension so the extremal surfaces appear as curves).  These surfaces foliate some portion of the bulk, and in an appropriate limit of the~$\lambda^i$ they degenerate to a point on~$\partial M$.}
\label{fig:foliation}
\end{figure}

First, let us list and motivate the important assumptions behind our argument.  Assume that there exists a bulk subregion~$\Rcal \subset M$ foliated by a~$(d-2)$-parameter family of two-dimensional surfaces~$\Sigma(\lambda^i)$,~$i = 3, \ldots, d$, with the following properties:
\begin{itemize}
	\item Each surface~$\Sigma(\lambda^i)$ is spacelike (with respect to~$g_{ab}$), topologically a disk, and is anchored to~$\partial M$ on a closed curve~$\partial \Sigma(\lambda^i)$;
	\item The foliation~$\{\Sigma(\lambda^i)\}$ contains a one-parameter subfamily of surfaces which converge to a point on~$\partial M$;
	\item Each~$\Sigma(\lambda^i)$ is extremal with respect to~$g_{ab}$;
	\item For each~$\lambda^i$,~$\Sigma(\lambda^i)$ is weakly stable in the sense of~\cite{EngFis19}: that is, for any curve~$\gamma \subset \partial M$ which is a sufficiently small perturbation of~$\partial \Sigma(\lambda^i)$, there exists an extremal surface anchored to~$\gamma$ which is a small perturbation of~$\Sigma(\lambda^i)$; and
	\item The area of the surface~$\Sigma(\lambda^i)$ is known, as is the area of any such sufficiently small extremal perturbation of it.
\end{itemize}
An illustration of this foliation is provided in Figure~\ref{fig:foliation}.  Also assume that the metric~$h_{ab}$ induced on~$\partial M$ from~$g_{ab}$ is known, as is the extrinsic curvature~$\Kcal_{ab}$ of~$\partial M$ in~$M$.  We will colloquially use the term ``boundary data'' to refer to the collection of objects consisting of the geometry~$(\partial M, h_{ab})$, curvature~$\Kcal_{ab}$, boundary curves~$\partial \Sigma(\lambda^i)$, and areas~$A[\Sigma(\lambda^i)]$ of the~$\Sigma(\lambda^i)$ and small (extremal) perturbations thereof.  Our main result is that under the above assumptions, this boundary data uniquely fixes the metric~$g_{ab}$ in~$\Rcal$ up to diffeomorphisms.

A reader may be concerned that because the geometries we typically consider in AdS/CFT are asymptotically locally AdS (AlAdS), the ``boundaries'' we are considering really correspond to a conformal structure at asymptotic infinity, rather than a finite boundary like~$\partial M$ as we consider here.  However, the process of holographic renormalization, which is by now quite well-understood~\cite{Skenderis:2002wp,Taylor:2016aoi}, is essentially the process of ``regulating'' the conformal boundary of an AlAdS spacetime by transforming it into a finite cutoff boundary\footnote{E.g., after introducing the Fefferman-Graham radial coordinate~$z$ associated to a particular choice of boundary conformal frame, we may consider the surface~$z = \eps > 0$ to be a regulated boundary and define regulated boundary data on it rather than on the bona fide conformal boundary at~$z = 0$.}.  The geometry~$(M, g_{ab})$ above should be interpreted as such a regulated AlAdS geometry, and its corresponding boundary data should be interpreted as the regulated version of UV-divergent objects.

The assumption that~$h_{ab}$ and~$\Kcal_{ab}$ are known is quite natural: the former is just the boundary metric (appropriately regulated), while~$\Kcal_{ab}$ is essentially the boundary stress tensor (up to known anomalies that depend on the intrinsic curvature of~$(\partial M, h_{ab})$, if~$d$ is odd).  Likewise, when~$d = 4$ (which is our case of primary interest), the areas of the~$\Sigma(\lambda^i)$ have a physical interpretation as well: as long as the~$\Sigma(\lambda^i)$ are the minimal-area extremal surfaces anchored to the boundary curves~$\partial \Sigma(\lambda^i)$, then the HRT proposal interprets their areas as (regularized) entanglement entropies of the boundary regions enclosed by the curves~$\partial \Sigma(\lambda^i)$.  For more general~$d$, the physical interpretation of these areas is less clear, but at least in certain contexts it is known that they correspond to the expectation values of Wilson loops in the dual CFT~\cite{Mal98}.  Consequently, what we refer to as boundary data are expected to correspond to objects in the dual CFT.

\begin{figure}[t]
\centering
\subfigure[]{
\label{subfig:bndryfoliation}
\includegraphics[page=2]{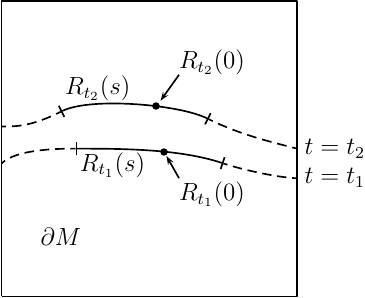}
}
\hspace{1cm}
\subfigure[]{
\label{subfig:bulkfoliation}
\includegraphics[page=3,width=0.35\textwidth]{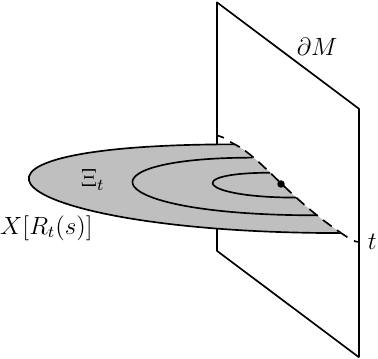}
}
\caption{In~$d = 4$, we sketch the construction of a family of extremal surfaces~$\Sigma(\lambda^i)$ which foliate~$\Rcal$ under some mild, but certainly nontrivial, assumptions.  \subref{subfig:bndryfoliation}: Consider some slicing of the boundary into slices of constant~$t \in (t_i, t_f)$, and on each such slice introduce a one-parameter family of regions~$R_t(s)$ which shrink to a point as~$s$ is decreased.  \subref{subfig:bulkfoliation}: for each~$t$, the HRT surfaces~$X[R_t(s)]$ will sweep out a three-dimensional surface~$\Xi_t$ (shaded) as long as they never jump.  If~$\Xi_t$ moves everywhere continuously to the future as~$t$ is increased, then the~$\Xi_t$ foliate some region~$\Rcal$, and the family of extremal surfaces~$\Sigma(t,s) = X[R_t(s)]$ do as well.}
\label{fig:EWNfoliation}
\end{figure}

How should we interpret the assumption that the~$\Sigma(\lambda^i)$ foliate~$\Rcal$?  Let us provide some intuition by giving a rough construction of such a family\footnote{We emphasize that the construction we are about to describe, which introduces a foliation of bulk time slices~$\Xi_t$, is simply meant to be illustrative; the existence of the slices~$\Xi_t$ is not necessary to obtain our result.}, again in the most relevant case of~$d = 4$.  Specifically, consider a time slicing of the boundary~$\partial M$ by some arbitrary time coordinate~$t$, and on each slice of constant~$t$, introduce a one-parameter family of regions~$R_t(s)$ such that as~$s$ decreases,~$R_t(s)$ shrinks into itself until it degenerates to a point at~$s = 0$, as shown in Figure~\ref{subfig:bndryfoliation}.  Now, entanglement wedge nesting --- a consequence of causality in the CFT --- requires that as~$s$ is decreased (and thus as~$R_t(s)$ shrinks), the corresponding HRT surface~$X[R_t(s)]$ must move in a spacelike direction towards~$R_t(s)$, as shown in Figure~\ref{subfig:bulkfoliation}.  Since~$R_t(s)$ eventually shrinks down to a point, then if the~$X[R_t(s)]$ change continuously under this shrinking (i.e.~if they never ``jump''), then the~$X[R_t(s)]$ sweep out an achronal three-dimensional surface~$\Xi_t$ which by construction is foliated by extremal surfaces.  Now if we assume that as the boundary time~$t$ is increased over a sufficiently small range of~$t \in (t_i, t_f)$,~$\Xi_t$ moves continuously towards the future, then the two-parameter family of surfaces~$\Sigma(t,s) = X[R_t(s)]$ obeys precisely the assumptions required above.  Specifically, since the~$\Xi_t$ foliate the neighborhood~$\Rcal = \cup_{t \in (t_i, t_f)} \Xi_t$, and since each~$\Xi_t$ is foliated by the~$X[R_t(s)]$, the~$\Sigma(t,s)$ must foliate~$\Rcal$.  Moreover, since by construction the~$\Xi_t$ degenerate to a point on~$\partial M$, the two-parameter family~$\Sigma(t,s)$ contains a one-parameter subfamily that does so as well.

The key assumptions in this construction were that the~$X[R_t(s)]$ change continuously and that the~$\Xi_t$ move uniformly to the future as~$t$ is increased.  There are of course many known cases of the former assumption being violated, most specifically in the context of entanglement shadows~\cite{Freivogel:2014lja}, but it is not too restrictive for our purposes: indeed, there are known cases of HRT surfaces entering the event horizon of dynamical black holes~\cite{Liu:2013iza,Liu:2013qca,Hubeny:2013dea}, provided that they do so not too late after the formation of the black hole.  Likewise, we expect that while the assumption that the~$\Xi_t$ move uniformly to the future as~$t$ is increased should not hold in full generality, it certainly holds in many known cases.  Therefore, while the assumption that the~$\Sigma(\lambda^i)$ foliate~$\Rcal$ is nontrivial, it is not so restrictive as to exclude cases of physical interest.

\subsection{Sketch of Argument}
\label{subsec:sketch}

In order to show that the boundary data listed above uniquely fix the metric~$g_{ab}$ in the bulk subregion~$\Rcal$, we mirror the structure of~\cite{AleBal17}, with modifications where necessary to deal with the fact that~$d \geq 4$.  This structure consists of four steps:
\begin{enumerate}
\item Fixing coordinates.  We first fix a unique coordinate system in~$\Rcal$, consisting of the foliation parameters~$\lambda^i$ as well as an isothermal coordinate system~$\{x^\alpha\}$ on each~$\Sigma(\lambda^i)$.
\item Fixing the normal inverse metric components.  We use the extremality condition on the $\Sigma(\lambda^i)$ to reduce the problem of fixing the inverse metric components~$g^{ij} = g^{ab} (d\lambda^i)_a (d\lambda^j)_b$ ``normal'' to the~$\Sigma(\lambda^i)$ to an inverse boundary value problem for an elliptic PDE, after which a uniqueness theorem gives us the desired result.
\item Fixing the off-diagonal inverse metric components.  By ``tilting'' the foliation~$\Sigma(\lambda^i)$ in different directions, we obtain an algebraic system of linear equations for the ``off-diagonal'' inverse metric components~$g^{\alpha i} = g^{ab} (dx^\alpha)_a (d\lambda^i)_b$, which fixes these components uniquely.
\item Fixing the conformal factor.  The remaining metric component corresponds to the induced metric on each surface~$\Sigma(\lambda^i)$. The condition that the~$\Sigma(\lambda^i)$ be extremal yields a first-order hyperbolic PDE for the conformal factor of the (conformally flat) metric on the~$\Sigma(\lambda^i)$, which again ensures uniqueness.
\end{enumerate}
The fact that the surfaces~$\Sigma(\lambda^i)$ are all extremal means that infinitesimal perturbations thereof are governed by an elliptic system of PDEs.  Specifically, given a one-parameter family~$\Sigma(s)$ of extremal surfaces, the deviation vector~$\eta^a \equiv (\partial_s)^a$ obeys the Jacobi equation
\be
\label{eq:formalJacobi}
J \eta^a_\perp = 0,
\ee
where~$\eta_\perp^a \equiv {P^a}_b \eta^b$ is the projection of~$\eta^a$ onto the normal bundle of~$\Sigma$.  The explicit form of the Jacobi operator~$J$ is given explicitly in~\eqref{eq:covariantJacobi} below, but importantly it is an elliptic differential operator on~$\Sigma$ that depends on its intrinsic and extrinsic geometry.  The derivation of~\eqref{eq:formalJacobi} can be found in the context of minimal surface theory in e.g.~\cite{ColMin}; in the context of perturbative dynamics of classical cosmic branes and strings in~\cite{Guv93,Car93,BatCar95}; and in the context of entanglement entropy in AdS/CFT in~\cite{Mos17,GhoMis17,LewPar18}; a broad review will be provided in~\cite{EngFis19}.  For the unfamiliar reader, we note that~\eqref{eq:formalJacobi} can be thought of as a generalization of the equation of geodesic deviation
\be
\label{eq:geodesicdeviation}
t^b \grad_b \left(t^c \grad_c \eta^a\right) + {R_{bcd}}^a t^b t^d \eta^c = 0
\ee
to higher-dimensional extremal surfaces (here~$t^a$ is the affinely-parametrized tangent to the geodesic).

In Section~\ref{subsec:bndrydata} we show that knowledge of the first and second variations of the areas of the~$\Sigma(\lambda^i)$ is sufficient to determine the Cauchy data of the operator~$J$: that is, given any perturbation of~$\partial \Sigma(\lambda^i)$ characterized by some deviation vector~$\eta^a$, second area variations determine the normal derivative~$N^b D_b \eta^a$ at~$\partial \Sigma(\lambda^i)$, where~$D_a$ is the covariant derivative on~$\Sigma(\lambda^i)$ and~$N^a$ is the unit normal to~$\partial \Sigma(\lambda^i)$ in~$\Sigma(\lambda^i)$.  With this observation made, we may now summarize each of the four steps listed above.

\subsubsection*{Fixing Coordinates}

To introduce the unique coordinate system on~$\Rcal$, note that since the extremal surfaces~$\Sigma(\lambda^i)$ are assumed to foliate~$\Rcal$, the parameters~$\lambda^i$ that label the members of this foliation provide a natural choice of~$(d-2)$ coordinates; moreover, since the boundary curves~$\partial \Sigma(\lambda^i)$ are known, these coordinates are uniquely specified by boundary data.  The remaining two coordinates, which we will denote in general as~$y^\alpha$~($\alpha = 1,2$), must be coordinates on each of the surfaces~$\Sigma(\lambda^i)$.  To fix these, we exploit the crucial fact that the~$\Sigma(\lambda^i)$ are two-dimensional, which implies that on each~$\Sigma(\lambda^i)$ there exist isothermal coordinates. We denote these isothermal coordinates by~$x^\alpha$, and the induced metric on~$\Sigma(\lambda^i)$ must then take the conformally flat form
\be
\label{eq:isothermal}
ds^2_\Sigma = e^{2\phi} \left[(dx^1)^2 + (dx^2)^2\right],
\ee
where~$\phi$ is a scalar on~$\Sigma(\lambda^i)$.  However, as we discuss in Section~\ref{subsec:isothermal}, because the isothermal coordinates~$\{x^\alpha\}$ of course depend on the induced metric on~$\Sigma(\lambda^i)$, there is no guarantee that we can choose to work in isothermal coordinates without spoiling the boundary data (in other words, two different metrics on~$\Sigma$ typically cannot both be brought to the form~\eqref{eq:isothermal} in the \textit{same} set of isothermal coordinates).  Fortunately, it turns out that the aforementioned fact that the Cauchy data of~$J$ is fixed by the boundary data ensures that this problem can be avoided.  To do so, the surface~$\Sigma(\lambda^i)$, which has finite boundary, is first artificially extended to an asymptotically flat manifold in a way that is fixed by boundary data.  Then a theorem of Ahlfors~\cite{Ahlfors} guarantees that there exists a unique set of isothermal coordinates~$\{x^\alpha\}$ on this asymptotically flat manifold which are fixed at infinity.  These unique coordinates range over the entire plane~$\mathbb{R}^2$, but the Cauchy data of~$J$ uniquely fixes the subregion of the plane that corresponds to the original (unextended) surface~$\Sigma(\lambda^i)$.  The restriction of these unique~$\{x^\alpha\}$ to this subregion thus yields a unique set of isothermal coordinates on each~$\Sigma(\lambda^i)$, and thus a unique coordinate chart~$\{x^\alpha, \lambda^i\}$.  We emphasize, however, that we only show the existence of the~$\{x^\alpha\}$; we do not give a construction of these coordinates from boundary data.  In the chart~$\{x^\alpha, \lambda^i\}$, the metric is naturally decomposed into three sets of components\footnote{It is slightly conceptually simpler to work with the inverse metric components for two reasons: first, the coordinate basis one-forms~$(d\lambda^i)_a$ are independent of the choice of coordinates on the~$\Sigma(\lambda^i)$, while the coordinate basis vectors~$(\partial_{\lambda^i})^a$ are not; second, the vectors~$(d\lambda^i)^a$ are normal to the~$\Sigma(\lambda^i)$, while in general the~$(\partial_{\lambda^i})^a$ need not be.}:
\be
g^{ij} \equiv g^{ab} (d\lambda^i)_a (d\lambda^j)_b, \qquad g^{\alpha i} \equiv g^{ab} (dx^\alpha)_a (d\lambda^i)_b, \qquad g^{\alpha\beta} \equiv g^{ab} (dx^\alpha)_a (dx^\beta)_b.
\ee
To prove uniqueness of the metric, we must prove uniqueness of the~$g^{ij}$, which we call the components\footnote{We also call them the normal components throughout this paper.} of the (inverse) metric on the normal bundle to the~$\Sigma(\lambda^i)$; of the~$g^{\alpha i}$, which we call the the off-diagonal components; and of the conformal factor~$\phi$ appearing in~\eqref{eq:isothermal}.

\subsubsection*{Fixing the normal metric components}

Once the existence of the~$\{x^\alpha\}$ is established, it is quite straightforward to show that the~$g^{ij}$ are unique.  We invoke a theorem of Albin, Guillarmou, Tzou, and Uhlmann~\cite{AlbGui13}, which implies that the Jacobi operator~$J$ is uniquely determined (up to gauge) by boundary data.  Since by construction the coordinate vectors~$(\partial_{\lambda^i})^a$ are deviation vectors along a family of extremal surfaces, they must all satisfy the Jacobi equation~\eqref{eq:formalJacobi}; we show in Section~\ref{subsec:normal} that this property uniquely fixes the~$g^{ij}$.  As for the isothermal coordinates~$\{x^\alpha\}$, here we only show that the~$g^{ij}$ are unique without providing a way of constructing them.

\subsubsection*{Fixing the off-diagonal metric components}

\begin{figure}[t]
\centering
\includegraphics[page=4,width=0.33\textwidth]{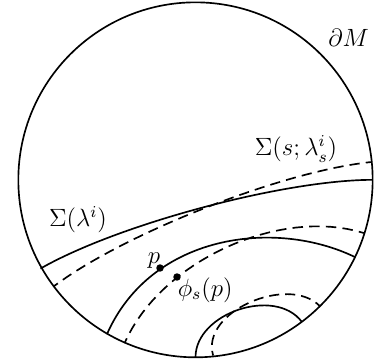}
\caption{To show uniqueness of the off-diagonal metric components, we deform the foliation~$\Sigma(\lambda^i)$ to a family of foliations~$\Sigma(s; \lambda_s^i)$.  This deformation is generated by a one-parameter group of diffeomorphisms~$\phi_s$ which are fixed by requiring that under the action of~$\phi_s$, the isothermal coordinates of each point remain unchanged.  In other words, if the point~$p$ is labeled by the isothermal coordinate values~$(x^1_*, x^2_*)$ on the surface~$\Sigma(\lambda^i = \lambda^i_*)$, then the mapped point~$\phi_s(p)$ must be labeled by the same values of the isothermal coordinates on the deformed surface~$\Sigma(s; \lambda_s^i = \lambda^i_*)$.}
\label{fig:shapedeformation}
\end{figure}

To show uniqueness of the~$g^{\alpha i}$, we consider globally perturbing the foliation~$\Sigma(\lambda^i)$ to a one-parameter family of foliations~$\Sigma(s;\lambda_s^i)$ parametrized by~$s$, as shown in Figure~\ref{fig:shapedeformation}.  Intuitively, one can think of this perturbation as ``tilting'' all the surfaces~$\Sigma(\lambda^i)$ in some continuous way; more precisely, this perturbation is generated by some diffeomorphism generated by~$\eta^a \equiv (\partial_s)^a$.  Since we have established that the normal components~$g^{ij}$ can be obtained from boundary data, so too must the normal components~$g_s^{ij} \equiv g^{ab}(d\lambda_s^i)_a (d\lambda_s^j)_b$ in this new foliation for each~$s$.  But as we show in Section~\ref{subsec:off-diag}, the~$g_s^{ij}$ are related in a known way to the unperturbed normal metric components~$g^{ij}$, the unperturbed off-diagonal metric components~$g^{\alpha i}$, and the generator~$\eta^a$ of the diffeomorphism that transforms the foliation~$\Sigma(\lambda^i)$ to the family~$\Sigma(s; \lambda_s^i)$.  Since this generator~$\eta^a$ is again a deviation vector field along a family of extremal surfaces, it too obeys the Jacobi equation~\eqref{eq:formalJacobi}, which as described above is known in terms of boundary data.  Consequently, we obtain an equation relating the known objects~$g^{ij}$,~$g^{ij}_s$,~$\eta^a_\perp$ to the unknown objects~$g^{\alpha i}$ and (the components of)~$\eta^a_\parallel \equiv {\sigma^a}_b \eta^b$.  By considering different ways of ``tilting'' the~$\Sigma(\lambda^i)$, we show that for~$d \geq 4$ it is possible to thereby obtain a set of linear algebraic equations for the~$g^{\alpha i}$ which can be solved in terms of known boundary data, thereby proving uniqueness of the off-diagonal metric components~$g^{\alpha i}$.

It is worth highlighting that taking~$d \geq 4$ is crucial, as for~$d = 3$ it is not possible to obtain enough independent linear algebraic equations to fix all of the~$g^{\alpha i}$; this is essentially due to the presence of the components of~$\eta^a_\parallel$, which are not fixed by the Jacobi equation.  Indeed, the~$d = 3$ case was dealt with in~\cite{AleBal17}, and requires deriving integral equations for the components of~$\eta^a_\parallel$, which in turn yield a system of \textit{integral} equations for the~$g^{\alpha i}$.  Using these equations to show that the~$g^{\alpha i}$ are uniquely fixed by boundary data requires making several quite strong additional assumptions on the geometry~$(M, g_{ab})$ which we are able to avoid here.

\subsubsection*{Fixing the conformal factor}

Finally, to show that the conformal factor~$\phi$ is unique, we show in Section~\ref{subsec:conformal} that the extremality condition~$K^a = 0$ reduces to a first-order linear PDE for~$\phi$ whose coefficients are all known functions of the (now uniquely fixed) metric components~$g^{ij}$,~$g^{\alpha i}$.  When restricted to a one-parameter subfamily of the~$\Sigma(\lambda^i)$ which converges to a point on~$\partial M$ (which exists by the assumptions of our argument), this PDE becomes a hyperbolic differential equation for~$\phi$ along this subfamily.  Since the value of~$\phi$ on the boundary~$\partial M$ is known, this equation can be evolved inwards from the boundary to obtain~$\phi$ everywhere along the three-dimensional surface~$\Xi$ foliated by the aforementioned one-parameter subfamily of the~$\Sigma(\lambda^i)$.  But since the~$\Sigma(\lambda^i)$ are a continuous foliation,~$\Xi$ can be deformed to run through any of the~$\Sigma(\lambda^i)$, leading us to conclude that~$\phi$ is uniquely fixed everywhere in~$\Rcal$.

\section{Detailed Argument}
\label{sec:details}

Let us now flesh out the sketch of the argument presented in the previous section.  We begin by briefly setting up some notation and conventions for the Jacobi operator~$J$ and the normal bundle and by taking stock of what information is easily accessible from boundary data, specifically from first and second variations of the area of extremal surfaces.  Then we provide the construction of the isothermal coordinates~$\{x^\alpha\}$ which allows us to show in turn uniqueness of the~$g^{ij}$,~$g^{i\alpha}$, and~$g^{\alpha\beta}$.

\subsection{The Jacobi Operator and the Normal Bundle}

Explicitly, the Jacobi equation~\eqref{eq:formalJacobi} governing the normal component~$\eta_\perp^a$ of the deviation vector along a family of extremal surfaces is
\begin{subequations}
\be
\label{eq:covariantJacobi}
0 = J \eta_\perp^a \equiv \Delta_\Sigma \eta_\perp^a + {Q^a}_b \eta_\perp^b,
\ee
where
\be
\Delta_\Sigma \eta_\perp^a \equiv {P^a}_b \sigma^{cd} \grad_c\left({P^b}_e {\sigma^f}_d \grad_f \eta_\perp^e\right)
\ee
is the Laplacian on the normal bundle of~$\Sigma$ and
\be
Q_{ab} \equiv {K_a}^{cd} K_{bcd} +  {P_a}^c {P_b}^d \sigma^{ef} R_{cedf}.
\ee
\end{subequations}
It will be useful to decompose the Jacobi operator~$J$ in a basis~$\{(n^i)_a\}$,~$i = 3, \ldots, d$ of the normal bundle of~$\Sigma$.  To do so, we define a covariant derivative~$\widehat{D}_a$ on the normal bundle of~$\Sigma$ as follows\footnote{The notation~$\widehat{D}_a$ is a bit redundant; the hat on~$\widehat{D}_a u^i$ is simply meant to emphasize that~$\widehat{D}_a u^i$ is the~$i$ component of the covariant derivative~${\sigma_a}^b \grad_b u^c$, which of course in general need not be the same as the covariant derivative of the \textit{scalar} components of~$u^a$.}: for any vector~$u^a$ in the normal bundle of~$\Sigma$,
\be
\label{eq:Dhatdef}
\widehat{D}_a u^i = (n^i)_c {\sigma_a}^b \grad_b u^c = D_a u^i - \sum_{j = 3}^d {\omega_{aj}}^i u^j,
\ee
where~$D_a$ is the usual covariant derivative on~$\Sigma$ compatible with the induced metric~$\sigma_{ab}$,~$u^i \equiv u \cdot n^i$ are the components of~$u^a$ in this basis, and the connection one-forms are given by
\be
\label{eq:omegadef}
{\omega_{ai}}^j = \sum_{k = 3}^d {\sigma_a}^b P_{ik} (n^k)^c \grad_b (n^j)_c,
\ee
where~$P_{ij}$ are the components of the matrix inverse of~$P^{ij} = n^i \cdot n^j$.  By construction, this covariant derivative is metric-compatible in the sense that~$\widehat{D}_a \sigma_{bc} = 0$,~$\widehat{D}_a P_{ij} = 0$, and likewise for~$\sigma^{ab}$ and~$P^{ij}$.  In terms of this derivative operator, we may write~$J$ as
\be
J \eta^i = -\widehat{D}^\dagger \widehat{D} \eta^i + \sum_{j = 3}^d {Q^i}_j \eta^j,
\ee
where~$\widehat{D}^\dagger$ is the (formal) adjoint of~$\widehat{D}$ under the inner product
\be
\label{eq:inprod}
\langle u|w\rangle_\sigma = \int_\Sigma P_{ij} u^i w^j \, \bm{\eps}_\sigma,
\ee
where~$\bm{\eps}_\sigma$ is the natural volume element on~$\Sigma$ constructed from the metric~$\sigma_{ab}$ (i.e.~in a coordinate system~$\{y^\alpha\}$ on~$\Sigma$,~$\bm{\eps}_\sigma = \sqrt{\sigma} \, dy^1 \wedge dy^2$, where~$\sigma$ is the determinant of the matrix of components~$\sigma_{\alpha\beta}$).  In particular, note that we have used the fact that~$(n^i)_a \Delta_\Sigma \eta^a_\perp = \sigma^{ab} \widehat{D}_a \widehat{D}_b \eta^i = -\widehat{D}^\dagger \widehat{D} \eta^i$.

\subsection{Boundary Data}
\label{subsec:bndrydata}

Let us now examine what information is extractable from the boundary data, which we remind the reader consists of the area of any spacelike two-dimensional extremal surface anchored to the boundary~$\partial M$; the boundary metric~$h_{ab}$ on~$\partial M$; the extrinsic curvature~$\mathcal{K}_{bc}$ of~$\partial M$ in~$M$; and the boundary surfaces~$\partial \Sigma(\lambda^i)$ (and hence also the parameters~$\{\lambda^i\}$ on the boundary).

Consider an extremal surface~$\Sigma$ anchored to~$\partial M$ and some deviation vector field~$\eta^a$ on it corresponding to a one-parameter family of boundary-anchored surfaces~$\Sigma(s)$; note that as all the~$\Sigma(s)$ are boundary-anchored,~$\eta^a$ must be tangent to~$\partial M$.  A standard result, presented in Appendix~\ref{app:areavariation}, is that the first variation of the area of these surfaces is just a boundary term:
\be
\label{eq:firstareavariation}
\left.\frac{dA[\Sigma(s)]}{ds} \right|_{s = 0} = \int_{\partial \Sigma} N_a \eta^a,
\ee
where~$N^a$ is the unit outward-pointing normal to~$\partial \Sigma$ in~$\Sigma$ (and we are leaving the natural volume form on~$\partial \Sigma$ implied).  Since we know the area of any extremal surface anchored to the boundary, we in particular know the area of any extremal surface anchored to an arbitrary deformation of~$\partial \Sigma$; this means we know the left-hand side of~\eqref{eq:firstareavariation} for \textit{any}~$\eta^a|_{\partial \Sigma}$ tangent to~$\partial M$.  Considering only deviation vectors~$\eta^a|_{\partial \Sigma}$ that have support on arbitrarily small portions of~$\partial \Sigma$ implies that we have access to the projection~${h_a}^b N_b$.  Since~$N^a$ is a unit vector, this means we can also obtain the inner product~$N \cdot v$, where~$v^a$ is the unit outward-pointing vector normal to~$\partial M$:
\be
\label{eq:Ndotv}
N \cdot v = \sqrt{1 - h_{ab} N^a N^b}.
\ee
Finally, we also show in Appendix~\ref{app:boundaryeta} that first area variations of each surface~$\Sigma(\lambda^i)$ are sufficient to recover the components~$g^{ij}$ of the inverse metric (in the coordinates~$\lambda^i$) at~$\partial \Sigma(\lambda^i)$.

Now consider a two-parameter family~$\Sigma(s_1, s_2)$ of boundary-anchored extremal surfaces with~$\Sigma(0,0) = \Sigma$; again we define the deviation vector fields~$\eta_1^a = (\partial_{s_1})^a|_{(s_1,s_2) = (0,0)}$,~$\eta_2^a = (\partial_{s_2})^a|_{(s_1,s_2) = (0,0)}$, and note that~$\eta_1^a|_{\partial \Sigma}$ and~$\eta_2^a|_{\partial \Sigma}$ are freely specifiable.  A first derivative simply reproduces~\eqref{eq:firstareavariation},
\be
\left.\frac{\partial A[\Sigma(s_1,s_2)]}{\partial s_1} \right|_{(s_1,s_2) = (0,0)} = \int_{\partial \Sigma} N_a \eta_1^a,
\ee
while as we show in Appendix~\ref{app:areavariation}, a second variation yields
\begin{multline}
\label{eq:secondareavariation}
\left.\frac{\partial^2 A[\Sigma(s_1,s_2)]}{\partial s_2 \, \partial s_1}\right|_{(s_1,s_2) = (0,0)} = \int_{\partial \Sigma} \left[\sum_{i = 3}^d \eta_1^i N^a \widehat{D}_a (\eta_2)_i \right. \\ \left. \phantom{\sum_{i = 1}^s} + N_a \eta_2^b \grad_b \eta_1^a + 2(N \cdot \eta_{(1}) (\eta_{2)})_a k^a - (N \cdot \eta_1) (N \cdot \eta_2) k^a N_a \right],
\end{multline}
where~$k^a$ is the mean curvature of~$\partial \Sigma$ in~$\partial M$.  The left-hand side of~\eqref{eq:secondareavariation} is of course known boundary data.  We claim that the second line of the right-hand side is known as well; this can be seen by noting that since~$\eta_1^a|_{\partial \Sigma}$,~$\eta_2^a|_{\partial \Sigma}$, and~$k^a$ are known and tangent to~$\partial M$, and since the projection of~$N^a$ onto~$\partial M$ is known from first area variations, the last two terms on the right-hand side of~\eqref{eq:secondareavariation} are known.  Moreover, since~$\eta^a_{1,2}|_{\partial \Sigma}$ are tangent to~$\partial M$, from the definition of extrinsic curvature we have
\be
N_a \eta_2^b \grad_b \eta_1^a = N_a \left[\eta_2^b \Dcal_b \eta_1^a + \eta_1^b \eta_2^c \Kcal^a_{\phantom{a}bc}\right] = N_a \eta_2^b \Dcal_b \eta_1^a + (N \cdot v) \, \eta_1^a \eta_2^b \Kcal_{ab},
\ee
where~$\Dcal_a$ is the covariant derivative on~$\partial M$ compatible with the boundary metric~$h_{ab}$, ${\Kcal^c}_{ab}$ is the extrinsic curvature of~$\partial M$ (and as is customary we wrote~$\Kcal_{ab} = v_c \Kcal^c_{ab}$), and we used that the component of~$N^a$ normal to~$\partial M$ is~$(N \cdot v) v^a$ (where as before,~$v^a$ is the outward-pointing unit normal to~$\partial M$).  But since~$\eta^a_{1,2}$ are tangent to~$\partial M$, so is~$\eta_2^a \Dcal_a \eta_1^b$, and thus its contraction with~$N^a$ is known from first area variations.  Likewise,~$N \cdot v$ is known from equation~\eqref{eq:Ndotv}, and~$\Kcal_{ab}$ is assumed known.  Thus the first term on the second line of~\eqref{eq:secondareavariation} is known as well.

We therefore conclude that the first term on the right-hand side of~\eqref{eq:secondareavariation} is known boundary data:
\be
\int_{\partial \Sigma} \sum_{i = 3}^d \eta_1^i N^a \widehat{D}_a (\eta_2)_i = \mbox{known}.
\ee
Now, the boundary values~$\eta_{1,2}^a|_{\partial \Sigma}$ are known boundary data and can be chosen arbitrarily; the normal bundle components~$\eta_{1,2}^i|_{\partial \Sigma}$, however, need not be known from boundary data, since they depend on the choice of basis~$\{(n^i)^a\}$, which in general will not be tangent to~$\partial M$.  Luckily, as we show in Appendix~\ref{app:boundaryeta}, when the basis~$\{(n^i)^a\}$ is chosen to be~$(n^i)^a = (d\lambda^i)^a$ (with the~$\lambda^i$ the parameters labeling the foliation~$\Sigma(\lambda^i)$ introduced above), the components~$\eta_{1,2}^i|_{\partial \Sigma}$ and~$(\eta_{1,2})_i|_{\partial \Sigma}$ can indeed be recovered from known boundary data thanks to the first area variation formula.  Then by considering~$\eta_1^i|_{\partial \Sigma}$ with support on arbitrarily small regions, we conclude that second area variations yield the Neumann boundary data~$N^a \widehat{D}_a (\eta_2)_i$ associated to any choice of~$(\eta_2)^i|_{\partial \Sigma}$.  But since~$\eta_2^a$ is the deviation vector along a family of extremal surfaces, it obeys the Jacobi equation~\eqref{eq:covariantJacobi}; thus knowledge of~$N^a \widehat{D}_a \eta^i_2$ for any~$\eta_2^a$ yields knowledge of the Cauchy data~$\Ccal_J$ of the Jacobi operator~$J$:
\be
\Ccal_J = \left\{ \left(\eta^i|_{\partial \Sigma}, N^a \widehat{D}_a \eta^i|_{\partial \Sigma}\right) \middle| J\eta^i = 0 \mbox{ on } \Sigma\right\}.
\label{eqn:cauchydata}
\ee
The main result of this section is thus that second variations of the area of a minimal surface~$\Sigma$ under arbitrary perturbations of its boundary fix~$\Ccal_J$, so~$\Ccal_J$ can be obtained from the boundary data.

\subsection{Fixing Coordinates on $\Sigma$}
\label{subsec:isothermal}

As discussed above, the parameters~$\lambda^i$ give~$(d-2)$ coordinates on~$M$ which are uniquely fixed by boundary data (i.e.~by the boundary curves~$\partial \Sigma(\lambda^i)$); the remaining two coordinates label individual points on the surfaces~$\Sigma (\lambda^i)$.  The purpose of this section is to show the existence of a set of isothermal coordinates in which the metric takes the isothermal form~\eqref{eq:isothermal} and in which the boundary data is known.

\begin{figure}[t]
\centering
\includegraphics[page=6,width=0.4\textwidth]{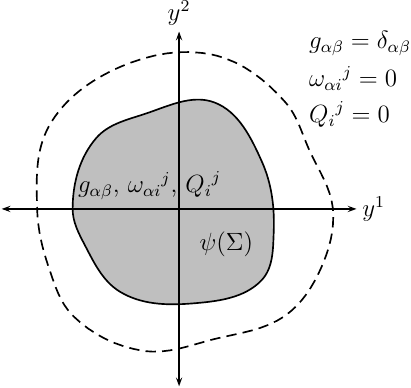}
\caption{The extension of~$\Sigma$ to an asymptotically flat manifold.  After choosing a coordinate system~$\{y^\alpha\}$ on~$\Sigma$ defined by the map~$\psi$, the metric components~$g_{\alpha\beta}$, connection one-forms~${\omega_{\alpha i}}^j$, and potential~${Q_i}^j$ are extended to the entire~$(y^1, y^2)$ plane by requiring that the former two vanish outside of~$\psi(\Sigma)$ while~$g_{\alpha\beta}$ should be continuous at~$\partial\psi(\Sigma)$ and equal to the Euclidean metric~$\delta_{\alpha\beta}$ outside of some set containing~$\psi(\Sigma)$, denoted by the dotted line.}
\label{fig:flatextension}
\end{figure}

First, let us choose some arbitrary set of coordinates~$\{y^\alpha\}$ on each~$\Sigma$ defined by a map~$\psi$ from~$\Sigma$ to some domain~$\psi(\Sigma) \subset \mathbb{R}^2$ of the~$(y^1,y^2)$ plane.  The map~$\psi$ is arbitrary, and in general there are three independent unknown metric components~$\sigma_{\alpha\beta} = g_{\alpha\beta}$ in this coordinate system.  A general set of isothermal coordinates~$\{x^\alpha\}$, in which the metric takes the form~\eqref{eq:isothermal}, is obtained by an additional map~$\Phi$ (which is not unique), as shown in Figure~\ref{fig:flatextension}.  Now, for any two metrics~$g_1$,~$g_2$ on~$\Sigma$, the uniformization theorem guarantees that the corresponding maps~$\Phi_1$,~$\Phi_2$ that put them in a conformally flat form~\eqref{eq:isothermal} can always be chosen so that the images~$\Phi_1(\psi(\Sigma))$,~$\Phi_2(\psi(\Sigma))$ in the~$(x^1, x^2)$ plane coincide.  But there is no guarantee that the maps will agree \textit{pointwise}, and in particular it need not be the case that for any~$p \in \partial \Sigma$,~$\Phi_1(\psi(p)) = \Phi_2(\psi(p))$; crucially, this implies that if the boundary data of~$g_1$ and~$g_2$ coincides on~$\Sigma$, it need \textit{not} coincide in the isothermal coordinates~$\{x^\alpha\}$ (by construction, it does coincide in the coordinates~$\{y^\alpha\}$, since they are defined by a single map~$\psi$).  However,~\cite{AleBal17} showed in their~$d = 3$ case that if (their version of) the Cauchy data~$\Ccal_{J_1}$,~$\Ccal_{J_2}$ agree on~$\Sigma$, then the maps~$\Phi_1$,~$\Phi_2$ \textit{can indeed} be chosen to agree pointwise on~$\partial \Sigma$, and thus preserve boundary data.  This result ensures that given any two metrics~$g_1$,~$g_2$ on~$\Sigma$ with matching Cauchy data, we can choose a gauge where both metrics take the form~\eqref{eq:isothermal} (on the same subset of~$\mathbb{R}^2$) while maintaining the matching of their boundary data.

\begin{figure}[t]
\centering
\includegraphics[page=5,width=0.9\textwidth]{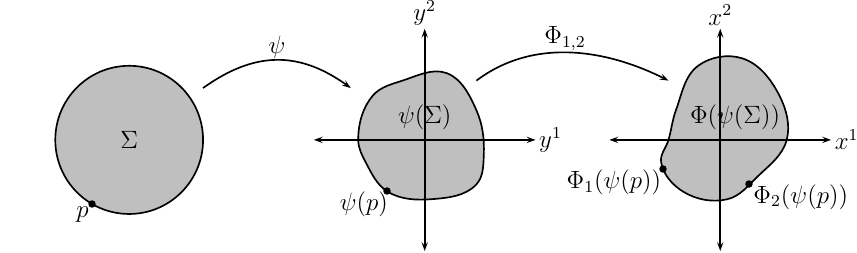}
\caption{A set of coordinates~$\{y^\alpha\}$ on~$\Sigma$ corresponds to a map~$\psi: \Sigma \to \mathbb{R}^2$.  A set of isothermal coordinates~$\{x^\alpha\}$ can be obtained by another map~$\Phi: \mathbb{R}^2 \to \mathbb{R}^2$.  For two different metrics~$g_1$,~$g_2$ on~$\Sigma$, the corresponding maps~$\Phi_1$,~$\Phi_2$ can be chosen to yield the same image~$\Phi_1(\psi(\Sigma)) = \Phi_2(\psi(\Sigma))$, but they need not agree pointwise; in other words, the isothermal coordinates of the point~$p$ obtained by the map~$\Phi_1$ need not be the same as those obtained by the map~$\Phi_2$, as shown.  Thus if the (pointwise) boundary data corresponding to~$g_1$ and~$g_2$ agrees on~$\Sigma$, it necessarily agrees in the coordinates~$\{y^\alpha\}$, but it need not agree in the coordinates~$\{x^\alpha\}$.}
\label{fig:coordinates}
\end{figure}

To generalize this result to our system, we proceed as follows.  First, we claim that given any two metrics~$g_1$,~$g_2$, there exist two bases~$\{(n^i_1)_a\}$,~$\{(n^i_2)_a\}$ of the normal bundle of~$\Sigma$ such that~(i) the components of these metrics in these bases agree, i.e.~$(g_1)^{ab} (n^i_1)_a (n^j_1)_b = (g_2)^{ab} (n^i_2)_a (n^j_2)_b \equiv \overline{P}^{ij}$; and~(ii) the Cauchy data~$\Ccal_{J_1}$ and~$\Ccal_{J_2}$ agree in this basis.  To see this, first note that in the coordinate basis~$(d\lambda^i)_a$, we have already established that~$\Ccal_{J_1} = \Ccal_{J_2}$ and that~$P^{ij}_1|_{\partial \Sigma} = P^{ij}_2|_{\partial \Sigma}$.  Now consider the basis transformation
\be
\label{eq:changebasis}
(n^i_2)_a \to \sum_{j = 3}^d {R^i}_j (n^j_2)_a
\ee
where we impose that~${R^i}_j|_{\partial \Sigma} = {\delta^i}_j$ and that
\be
\sum_{k,n = 3}^d {R^i}_k {R^j}_n P^{kn}_2 = P^{ij}_1;
\ee
this latter condition can always be satisfied since~$P^{ij}_1$ and~$P^{ij}_2$ are invertible.  Under such a transformation we have that~$P^{ij}_2 \to P^{ij}_1$, verifying claim~(i), while since~${R^i}_j$ is the identity at~$\partial \Sigma$, the boundary data is unaffected, verifying claim~(ii).  In fact, note that we may always perform an additional change of basis to make~$\overline{P}_{ij}$ constant on~$\Sigma$; for convenience, we will work in such a basis in what follows, though this simplification isn't strictly necessary.  In such a basis, it follows (either from the metric-compatibility condition~$\widehat{D}_a \overline{P}^{ij} = 0$ or directly from the definition~\eqref{eq:omegadef}) that~${\omega_a}^{(ij)} = 0$ (where normal bundle indices are raised and lowered with~$\overline{P}^{ij}$ and~$\overline{P}_{ij}$).

Now we make a further claim: for any metric and for a given~$\overline{P}^{ij}$, we can choose a basis in which the connection one-forms~${\omega_{a i}}^j$ obey~$N^a {\omega_{a i}}^j = 0$ at~$\partial \Sigma$.  To see this, again perform a change of basis~\eqref{eq:changebasis}, except now require that~${R^i}_j$ satisfy
\bea
{R^i}_j|_{\partial \Sigma} &= {\delta^i}_j, \\
N^a \partial_a {R^i}_j|_{\partial \Sigma} &= -N^a {\omega_{a j}}^i|_{\partial \Sigma}, \\
\sum_{k,n = 3}^d {R^i}_k {R^j}_n \overline{P}^{kn} &= \overline{P}^{ij}.
\eea
That these conditions are compatible can roughly be seen from a counting argument. The latter condition only places~$(d-2)(d-1)/2$ constraints on the~$(d-2)^2$ components of~${R^i}_j$ (since~$\overline{P}^{ij}$ is symmetric), while the second condition only constrains (the derivatives of)~$(d-2)(d-3)/2$ components of~${R^i}_j$ at~$\partial \Sigma$ (since~${\omega_a}^{ij}$ is antisymmetric).  Indeed, it is possible to check explicitly that the second constraint implies that
\be
\sum_{k,n = 3}^d  N^a \partial_a \left({R^i}_k {R^j}_n \overline{P}^{kn}\right)|_{\partial \Sigma} = 0,
\ee
which is consistent with the third.  By construction, such a transformation leaves boundary data invariant (specifically, it leaves~$C_J$ unchanged), while it is straightforward to check that~$N^a {\omega_{ai}}^j|_{\partial \Sigma} \to 0$, as claimed.  The upshot is therefore that for any two metrics~$g_1$,~$g_2$ on~$M$ with matching boundary data, we can find (generally different) bases on the normal bundle of~$\Sigma$ in which~$P^{ij}_1 = P^{ij}_2 = \overline{P}^{ij}$,~$N^a {(\omega_1)_{ai}}^j|_{\partial \Sigma} = N^a {(\omega_2)_{ai}}^j|_{\partial \Sigma} = 0$, and~$\Ccal_{J_1} = \Ccal_{J_2}$.

Let us now work in such a basis in the arbitrary coordinate system~$\{y^\alpha\}$ on~$\Sigma$.  The metric components~$g_{\alpha\beta}$, the components~${\omega_{\alpha i}}^j$ of the connection one-forms, and the potential~${Q_i}^j$ are of course only defined in the image~$\psi(\Sigma) \subset \mathbb{R}^2$ of~$\Sigma$ in this coordinate chart.  However, let us now extend all of these objects to the entire~$(y^1, y^2)$ plane by taking~${Q_i}^j = 0$ and~${\omega_{\alpha i}}^j = 0$ outside of~$\psi(\Sigma)$, as well as taking the~$g_{\alpha\beta} = \delta_{\alpha\beta}$ outside of some set containing~$\psi(\Sigma)$ with~$g_{\alpha\beta}$ taken to be continuous at~$\partial \psi(\Sigma)$, as shown in Figure~\ref{fig:flatextension}.  Note that the value of the~$g_{\alpha\beta}$ at~$\partial \psi(\Sigma)$ is fixed by boundary data, since the line element along~$\partial \Sigma$ is\footnote{This can perhaps be seen most explicitly by taking the~$\{y^\alpha\}$ to be Gaussian normal coordinates near~$\partial \Sigma$.  In fact, in these coordinates it's clear that knowledge of the extrinsic curvature~$\Kcal_{ab}$ of~$\partial M$ in~$M$ would be sufficient to fix the derivatives~$N^a \partial_a g_{\alpha\beta}$ at~$\partial \psi(\Sigma)$ as well.}; thus the extension of~$g_{\alpha\beta}$ outside of~$\psi(\Sigma)$ can be chosen knowing only the boundary data corresponding to the metric~$g_{ab}$.

Having performed this extension to~$\mathbb{R}^2$, consider the ``exterior'' boundary-value problem
\begin{subequations}
\label{eqs:exteriorbndryproblem}
\begin{align}
J \eta^j_\xi = 0 \mbox{ on } \Omega_y \equiv \mathbb{R}^2 \setminus \psi(\Sigma), \\
e^{-(y^1 + i y^2)\xi} \eta^j_\xi(y) - \bar{\eta}^j \to 0 \mbox{ at large } y^\alpha, \\
\left(\eta^j_\xi|_{\partial \psi(\Sigma)}, N^a \widehat{D}_a \eta^j_\xi|_{\partial \psi(\Sigma)}\right) \in \Ccal_J,
\end{align}
\end{subequations}
where~$\bar{\eta}^j$ are arbitrary fixed nonzero numbers,~$\xi$ is an arbitrary nonzero complex number, and~$\Ccal_J$ is the Cauchy data of~$J$ in~$\psi(\Sigma)$.  We now argue that this problem has a unique solution for any~$\xi$ and~$\bar{\eta}^j$.  To do so, first consider the problem
\begin{subequations}
\label{eqs:bndryproblem}
\begin{align}
J \eta^j_\xi = 0 \mbox{ on } \mathbb{R}^2, \\
e^{-(y^1 + i y^2)\xi} \eta^j_\xi(y) - \bar{\eta}^j \to 0 \mbox{ at large } y^\alpha.
\end{align}
\end{subequations}
The Jacobi operator can be expanded in terms of the connection coefficients as
\begin{multline}
\label{eq:Jexpansion}
J \eta^i = \sum_{\alpha, \beta = 1}^2 \left\{ \frac{1}{\sqrt{\sigma}} \partial_\alpha \left(\sqrt{\sigma} \sigma^{\alpha\beta} \partial_\beta \eta^i \right) \phantom{\sum_{j=3}^d} \right. \\ \left. + \sigma^{\alpha\beta} \sum_{j = 3}^d \left[-2 {\omega_{\alpha j}}^i D_\beta \eta^j - (D_\alpha {\omega_{\beta j}}^i) \eta^j + \sum_{k = 1}^d {\omega_{\alpha j}}^i {\omega_{\beta k}}^j \eta^k\right] \right\} + \sum_{j = 3}^d {Q^i}_j \eta^j,
\end{multline}
where~$\sigma^{\alpha\beta}$ is the matrix inverse of~$\sigma_{\alpha\beta} = g_{\alpha\beta}$ and~$\sigma$ is the determinant of~$\sigma_{\alpha\beta}$.  From the fact that~$g_{\alpha\beta}$ and~$N^a {\omega_{a i}}^j$ are continuous across~$\partial \psi(\Sigma)$ (by construction), the coefficients in the differential operator may be discontinuous at~$\partial \psi(\Sigma)$, but they are finite (i.e.~no derivatives of discontinuous objects appear in~\eqref{eq:Jexpansion}).  Solutions to~\eqref{eqs:bndryproblem} must therefore be differentiable, and since by construction any solution to~\eqref{eqs:bndryproblem} has Cauchy data in~$\Ccal_J$ at~$\partial \psi(\Sigma)$, any solution to~\eqref{eqs:bndryproblem} is therefore a solution to~\eqref{eqs:exteriorbndryproblem}.  We therefore look for solutions to~\eqref{eqs:bndryproblem}.

To do so, convert to isothermal coordinates~$\{x^\alpha\}$ via a map~$\Phi: \mathbb{R}^2 \to \mathbb{R}^2$, where note that now~$\Phi$ is interpreted as mapping the entire~$(y^1, y^2)$ plane to the entire~$(x^1, x^2)$ plane in such a way that the extended metric takes the form~\eqref{eq:isothermal} everywhere.  In fact, a theorem of Ahlfors~\cite{Ahlfors} guarantees that there exists a \textit{unique} such set of coordinates with the property that~$x^\alpha(y) \to y^\alpha$ at large~$y^\alpha$.  In these coordinates, the Laplacian on the normal bundle is just~$\widehat{D}^2 \eta^i = e^{-2\phi} \widehat{D}^2_{g_E} \eta^i$, where~$\widehat{D}^2_{g_E}$ is the Laplacian on the normal bundle with respect to the flat metric~$(g_E)_{\alpha\beta} = \delta_{\alpha\beta}$.  Consequently, in these new coordinates the boundary-value problem~\eqref{eqs:bndryproblem} becomes
\begin{subequations}
\label{eqs:isothermalbndryproblem}
\begin{align}
e^{2\phi} J \eta^j_\xi = \widehat{D}^2_{g_E} \eta^j_\xi + \sum_{k = 3}^d e^{2\phi} {Q^j}_k \eta^k_\xi = 0 \mbox{ on } \mathbb{R}^2, \label{subeq:isothermalJ} \\
e^{-(x^1 + i x^2)\xi} \eta^j_\xi(x) - \bar{\eta}^j \to 0 \mbox{ at large } x^\alpha.\label{subeq:isothermalbndrycond}
\end{align}
\end{subequations}
To obtain solutions to these equations, let us define
\be
\delta \eta^j_\xi(x) \equiv e^{-(x^1 + i x^2)\xi} \eta^j_\xi(x) - \bar{\eta}^j;
\ee
Given this, then the boundary condition~\eqref{subeq:isothermalbndrycond} is clearly the requirement that~$\delta \eta^j_\xi(x)$ vanish at large~$x^\alpha$, while the Jacobi equation~\eqref{subeq:isothermalJ} becomes
\be
\label{eq:Fdeltaeta}
F_\xi \delta \eta^j_\xi = -F_\xi \bar{\eta}^i,
\ee
where we have defined the operator
\bea
F_\xi u^j &\equiv e^{2\phi} e^{-(x^1 + i x^2)\xi} J \left(e^{(x^1 + i x^2)\xi} u^j \right), \\
		&= \widehat{D}^2_{g_E} u^j + \sum_{k = 3}^d e^{2\phi} {Q^j}_k u^k + \xi\left[\left(\partial_1 + i \, \partial_2\right) u^j - 2 \sum_{k = 3}^d \left({\omega_{1k}}^j + i \, {\omega_{2k}}^j\right) u^k\right].
\eea
Note in particular that~$F$ is a uniformly elliptic operator, and that the right-hand side of~\eqref{eq:Fdeltaeta} vanishes in the exterior region~$\Omega_x \equiv \Phi(\Omega_y)$ (since~$\bar{\eta}^j$ is constant and both~${Q^i}_j$ and~${\omega_{a i}}^j$ vanish there).  The equation~\eqref{eq:Fdeltaeta} with boundary condition~$\delta \eta^i(x) \to 0$ at large~$x^\alpha$ is therefore an elliptic Dirichlet problem, which can be solved uniquely (assuming nondegeneracy of~$F_\xi$, which follows from the nondegeneracy of~$J$) by integrating against the Dirichlet Green's function of~$F_\xi$:
\be
\label{eq:deltaetasolution}
\delta \eta^j_\xi(x) = -\int \sum_{k = 3}^d {G^j}_k(x,x') F_\xi \bar{\eta}^k(x') \, d^2 x',
\ee
where the~${G^j}_k(x,x')$ satisfy
\begin{multline}
\widehat{D}^2_{g_E} {G^j}_k(x,x') + \sum_{n = 3}^d e^{2\phi(x)} {Q^j}_n(x) {G^n}_k(x,x') \\ + \xi\left[\left(\partial_1 + i \, \partial_2\right) {G^j}_k(x,x') - 2 \sum_{n = 3}^d \left({\omega_{1n}}^j(x) + i \, {\omega_{2n}}^j(x)\right) {G^j}_k(x,x')\right] = \delta(x,x') {\delta^j}_k
\end{multline}
with all derivatives acting on the first argument of~${G^j}_k(x,x')$, and with~${G^j}_k(x,x')$ vanishing at large~$x - x'$.  Note that here we are assuming the existence and uniqueness of~${G^j}_k(x,x')$, which seems quite reasonable to us on physical grounds; nevertheless, it is for this reason that we call our result an argument rather than a proof.  Proceeding under this assumption,~\eqref{eq:deltaetasolution} shows existence and uniqueness of a solution to~\eqref{eqs:isothermalbndryproblem}, and therefore to~\eqref{eqs:exteriorbndryproblem}.  In fact, because the operator~$F_\xi$ is a flat-space Laplacian plus lower-derivative correction terms, we expect that the asymptotic falloff of~${G^j}_k(x,x')$ should be exponential; such a falloff is more than sufficient to ensure that for each~$j = 3, \ldots, d$, the norm\footnote{The asymptotic statements made in~\cite{AleBal17} actually used an~$L^2_{-\delta}$ norm; our argument here is not sufficiently precise to make this distinction relevant, so we just use the usual~$L^2$ norm for simplicity.}
\be
\left\| \delta \eta^j_\xi \right\|_{L^2(\Omega_x)} = \left\| e^{-(x^1 + i x^2)\xi} \eta^j_\xi(x) - \bar{\eta}^j \right\|_{L^2(\Omega_x)}
\ee
exists (where the norm is the usual~$L^2$ norm taken over the exterior region~$\Omega_x$ in the~$(x^1, x^2)$ plane).  Moreover, this norm vanishes as~$|\xi| \to \infty$, which can be seen by noting that in the exterior region~$\Omega_x$, the equation~\eqref{eq:Fdeltaeta} becomes simply
\be
\label{eq:largexi}
D^2_{g_E} \delta \eta^j_\xi + \xi(\partial_1 + i \partial_2) \delta \eta^j_\xi = 0,
\ee
with~$D^2_{g_E} = \partial_1^2 + \partial_2^2$ just the usual flat-space Laplacian.  In the limit~$|\xi| \to \infty$, this equation is dominated by the first-derivative terms, which require that~$\partial_\alpha \delta \eta^j_\xi \to 0$\footnote{In principle each derivative could instead be~$\Ocal(\xi)$, so each term in~\eqref{eq:largexi} would be~$\Ocal(\xi^2)$ thereby allowing for some cancellation, but we expect that the Dirichlet boundary conditions on~$\delta \eta^i_\xi$ must impose that the (unique) solution~$\delta \eta^i_\xi$ should have vanishing derivatives at large~$x^\alpha$, excluding such behavior.}.  But since~$\delta \eta^j_\xi \to 0$ asymptotically, the vanishing of its derivatives implies the vanishing of~$\delta \eta^j_\xi$ as well.  We therefore conclude that
\be
\label{eq:xinorm}
\left\| e^{-(x^1 + i x^2)\xi} \eta^j_\xi(x) - \bar{\eta}^j \right\|_{L^2(\Omega_x)} \to 0 \mbox{ as } |\xi| \to \infty.
\ee
It is worth pausing to make a brief remark on the interpretation of this statement.  In the exterior region~$\Omega_x$,~\eqref{subeq:isothermalJ} are just~$(d-2)$ decoupled Laplace equations, which can all be solved subject to the boundary condition~\eqref{subeq:isothermalbndrycond} by simply setting~$\eta^j_\xi(x) = \bar{\eta}^j e^{(x^1 + i x^2)\xi}$ everywhere.  However, the presence of the potential and connection coefficients in the interior region~$\Phi (\psi(\Sigma))$ perturbs these Laplace equations in some nontrivial way, so the solution to the problem~\eqref{eqs:isothermalbndryproblem} on the whole plane is corrected away from the pure exponential form~$\bar{\eta}^j e^{(x^1 + i x^2)\xi}$; the object~$\delta \eta^j_\xi$ is precisely this correction.  The behavior~\eqref{eq:xinorm} is simply the statement that as~$|\xi|$ grows, the asymptotic exponential behavior~$e^{(x^1 + i x^2)\xi}$ becomes sufficiently dominant over any perturbations in~$\Phi(\psi(\Sigma))$ that the correction~$\delta \eta^j_\xi$ vanishes in~$\Omega_x$.

We have therefore argued that the exterior boundary-value problem~\eqref{eqs:exteriorbndryproblem} has a unique solution for each~$\xi$, and moreover from~\eqref{eq:xinorm} we have that in terms of the~$L^2$ norm in the exterior region~$\Omega_y$ of the~$(y^1, y^2)$ plane, this solution obeys (for each~$j$)
\be
\label{eq:xinormy}
\left\| e^{-(x^1(y) + i x^2(y))\xi} \eta^j_\xi(y) - \bar{\eta}^j \right\|_{L^2(\Omega_y)} \to 0 \mbox{ as } |\xi| \to \infty,
\ee
where~$x^\alpha(y)$ are the unique isothermal coordinates introduced above.  As in the proof of Proposition~2.3 of~\cite{AleBal17}, we may now use this result to conclude that given two metrics~$g_1$,~$g_2$ with the same boundary data, we may find a shared set of isothermal coordinates~$x^\alpha$ in which both metrics take the form~\eqref{eq:isothermal} without spoiling the agreement of their boundary data.  To do so, first put both metrics (and the correpsonding connection coefficients and potentials) into the shared~$\{y^\alpha\}$ coordinates via the map~$\phi$, and extend both of these metrics to all of~$\mathbb{R}^2$ as described above.  Because by construction this extension is the same for both metrics, and because the Cauchy data of these metric is assumed to agree (in the~$\{y^\alpha\}$ coordinates), the exterior problem~\eqref{eqs:exteriorbndryproblem} is the same for both metrics as well.  Since we argued that the solution to the exterior boundary problem is unique, it must be the same for both metrics; we denote this solution by~$\eta_\xi^j$.  Now consider the two sets~$\{x_1^\alpha\}$,~$\{x_2^\alpha\}$ of unique isothermal coordinates corresponding to these two metrics.  The asymptotic behavior~\eqref{eq:xinormy} must hold in both of these coordinates, and therefore by the triangle inequality we must also have for each~$j$
\begin{subequations}
\be
\left\|e^{-(x_1^1(y) + i x_1^2(y))\xi} \eta^j_\xi(y) - e^{-(x_2^1(y) + i x_2^2(y))\xi} \eta^j_\xi(y) \right\|_{L^2(\Omega_y)} \to 0 \mbox{ as } |\xi| \to \infty,
\ee
or
\be
\label{eq:differencebound}
\left\|\left(e^{-(\Delta x^1(y) + i \Delta x^2(y))\xi} - 1\right) e^{-(x_2^1(y) + i x_2^2(y))\xi} \eta^j_\xi(y)\right\|_{L^2(\Omega_y)} \to 0 \mbox{ as } |\xi| \to \infty,
\ee
\end{subequations}
where we have defined~$\Delta x^\alpha(y) \equiv x_1^\alpha(y) - x_2^\alpha(y)$.  Now proceed by contradiction: assume that~$\Delta x^1(y) > 0$ at some point~$y_0 \in \Omega_y$.  This implies by continuity that~$\Delta x^1(y) > 0$ in some neighborhood of~$y_0$; thus taking~$\xi = -c$ for some real~$c$, we have
\be
\left|e^{-(\Delta x^1(y) + i \Delta x^2(y))\xi} - 1\right| = \left|e^{(\Delta x^1(y) + i \Delta x^2(y))c} - 1\right| \to \infty \mbox{ as } c \to \infty.
\ee
This clearly violates the behavior~\eqref{eq:differencebound}.  Analogous arguments therefore imply that~$\Delta x^\alpha(y) = 0$ for all~$y \in \Omega_y$, thus establishing that the isothermal coordinate systems~$x_1^\alpha$,~$x_2^\alpha$ must agree everywhere in the exterior region~$\Omega_y$.  In particular, this includes the boundary~$\partial \Omega_y = \partial \psi(\Sigma)$.  (It is for this reason that~$\Sigma$ was taken to be a topological disk: if it were not, then~$\Omega_y$ would have more than one connected component, and we would only be able to conclude that the~$x_1^\alpha(y)$ and~$x_2^\alpha(y)$ agree in the component of~$\Omega_y$ containing the asymptotic region of the~$\mathbb{R}^2$.)

As desired, we consequently find that there exists a set of isothermal coordinates on~$\Sigma$ in which both metrics~$g_1$,~$g_2$ take the form~\eqref{eq:isothermal} and in which their boundary data agree.  The upshot, and the key result of this section, is therefore that there exists a preferred coordinate system~$\{x^\alpha\}$ on~$\Sigma$ which is fixed only by boundary data, and therefore we can endow all of~$M$ with a unique coordinate system~$\{x^\alpha, \lambda^i\}$ in which the induced metric on each~$\Sigma(\lambda^i)$ takes the form~\eqref{eq:isothermal}.  Moreover, the conformal factors~$\phi_1$,~$\phi_2$ of the corresponding metrics must agree at~$\partial \Phi(\psi(\Sigma))$, since they are obtained from the metric components~$g_{\alpha\beta}|_{\partial \psi(\Sigma)}$ in the~$\{y^\alpha\}$ coordinate system, which as we discussed above must match.

For future reference, let us note some useful results regarding the relationship between the metric components~$g_{\mu\nu}$ and inverse metric components~$g^{\mu\nu}$ in these preferred coordinates.  To do so, let us write the components~$g_{\mu\nu}$ as a matrix in block-diagonal form:
\be
g = \begin{pmatrix} e^{2\phi} I & B \\ B^T & C \end{pmatrix},
\ee
where~$I$ is the~$2 \times 2$ identity matrix,~$B$ is the~$2 \times (d-2)$ matrix with components~$g_{\alpha i}$, and~$C$ is the~$(d-2) \times (d-2)$ matrix with components~$g_{ij}$.  The blockwise inversion of~$g_{\mu\nu}$ then yields
\begin{subequations}
\label{eqs:ginverse}
\begin{align}
g^{-1} &= \begin{pmatrix} e^{-2\phi} I + e^{-4\phi} B (C-e^{-2\phi} B^T B)^{-1} B^T & -e^{-2\phi} B (C-e^{-2\phi} B^T B)^{-1} \\ - e^{-2\phi}  (C-e^{-2\phi} B^T B)^{-1}B^T & (C-e^{-2\phi} B^T B)^{-1} \end{pmatrix}, \\
	&\equiv \begin{pmatrix} \widetilde{A} & \widetilde{B} \\ \widetilde{B}^T & \widetilde{C} \end{pmatrix},
\end{align}
\end{subequations}
where we have defined~$\widetilde{A}$ as the~$2 \times 2$ matrix with components~$g^{\alpha\beta}$,~$\widetilde{B}$ as the~$2 \times (d-2)$ matrix with components~$g^{\alpha i}$, and~$\widetilde{C}$ as the~$(d-2) \times (d-2)$ matrix with components~$g^{ij}$.  We now immediately notice two features: (i)~$\widetilde{C}$ is manifestly invertible, and (ii)~$\widetilde{B} = -e^{-2\phi} B \widetilde{C}$, and therefore~$B = - e^{2\phi} \widetilde{B} \widetilde{C}^{-1}$.  This latter statement implies that the metric components~$g_{\alpha i}$ are related to the inverse metric components~$g^{ij}$,~$g^{\alpha i}$ as
\be
\label{eq:gialpha}
g_{\alpha i} = e^{2\phi} f_{\alpha i} (g^{\alpha i},g^{ij}),
\ee
where~$f_{\alpha i}$ are known functions.  This observation will later be useful in proving uniqueness of the conformal factor~$\phi$.

\subsection{The $g^{ij}$}
\label{subsec:normal}

Having established the existence of the preferred coordinate system~$\{x^\alpha, \lambda^i\}$, we may begin to prove the uniqueness of the metric by proving uniqueness of its components in this coordinate system.  We begin by proving uniqueness of the normal metric components~$g^{ij}$, which relies crucially on a theorem of~\cite{AlbGui13}.

To proceed, consider two metrics~$g_A$,~$A = 1,2$ on~$M$, and consider a particular slice~$\Sigma$ of the foliation~$\Sigma(\lambda^i)$ and work in the shared isothermal coordinates~$\{x^\alpha\}$ on that slice.  As discussed in Section~\ref{subsec:bndrydata}, let us also introduce two bases~$\{(n^i_A)_a\}$ of the normal bundle of~$\Sigma$ such that~$(n_1^i)_a|_{\partial \Sigma} = (n_2^i)_a|_{\partial \Sigma}$ and in which the components of the inverse metric are equal:~$(g_1)^{ab} (n_1^i)_a (n_1^j)_b = (g_2)^{ab} (n_2^i)_a (n_2^j)_b \equiv P^{ij}$ (it's worth nothing that we do not decorate~$P^{ij}$ with an overline because here it need not be constant, in contrast with the previous section).  Now, in the isothermal coordinates, the respective Jacobi operators~$J_A$ are given by
\be
J_A \eta^i = -e^{-2\phi_A} (\widehat{D}_{g_E})_A^\dagger \widehat{D}_A\eta^{i} + \sum_{j = 3}^d {(Q_A)^i}_j \eta^j,
\ee
%J_A \eta^i = -e^{-2\phi_A} (\widehat{D}_{g_E})_A^\dagger \widehat{D}_A + \sum_{j = 3}^d {(Q_A)^i}_j \eta^j,
where~$\phi_A$ are the conformal factors of~$g_A$ in the coordinates~\eqref{eq:isothermal},~$\widehat{D}_A$ is the covariant derivative defined as in~\eqref{eq:Dhatdef} with connection coefficients~${(\omega_A)_{\alpha i}}^j$, and~$(\widehat{D}_{g_E})_A^\dagger$ is the adjoint of~$\widehat{D}_A$ with respect to the inner product~\eqref{eq:inprod} with the induced metric on~$\Sigma$ just the flat metric,~$\sigma_{\alpha\beta} = \delta_{\alpha\beta}$, and the metric on the normal bundle~$P^{ij}$.  We emphasize that this inner product is the same for both metrics~$g_A$.  Now, consider the conformally rescaled Jacobi operator
\be
\label{eq:conformalJacobi}
\widetilde{J}_A \eta^i \equiv e^{2\phi_A} J_A \eta^i = -(\widehat{D}_{g_E})_A^\dagger \widehat{D}_A \eta^{i}+ \sum_{j = 3}^d e^{2\phi_A} {(Q_A)^i}_j \eta^j;
\ee
it is easy to see from the conformal transformation properties of the normal vector~$N^a$ that the Cauchy data of~$J_A$ is the same as the Cauchy data of~$\widetilde{J}_A$.  Since the Cauchy data of~$J_A$ agree, i.e.~$\Ccal_{J_1} = \Ccal_{J_2}$, we therefore also have that the Cauchy data of~$\widetilde{J}_A$ agree.

Now we invoke the theorem of~\cite{AlbGui13}: since the operators~$\widetilde{J}_1$ and~$\widetilde{J}_2$ have the same Cauchy data, they are the same up to gauge\footnote{Some caveats are in order: strictly speaking, the theorem of~\cite{AlbGui13} says to consider two operators~$D^\dagger_1 D_1 + V_1$ and~$D^\dagger_2 D_2 + V_2$ acting on a complex vector bundle with Hermitian inner product over a Riemann surface; if these operators have the same Cauchy data, then the connections~$D_1$ and~$D_2$ and the potentials~$V_1$ and~$V_2$ are related by a gauge transformation.  The vector bundle in our case is the normal bundle of~$\Sigma$, which is real, not complex; this does not affect the results of~\cite{AlbGui13}.  Perhaps more concerningly, the inner product~\eqref{eq:inprod} we use is not Hermitian due to the indefinite sign of~$P^{ij}$ (since we are working in a Lorentzian setting); fortunately, this lack of Hermiticity also does not affect the proof of the theorem~\cite{Gunther}.}.  That is, there must exist some~${R^i}_j$ such that
\begin{subequations}
\label{eqs:Gunther}
\begin{align}
{R^i}_j|_{\partial \Sigma} &= {\delta^i}_j, \\
P^{ij} &= \sum_{k,n = 3}^d {R^i}_k {R^j}_n P^{kn}, \label{subeq:Pijconserved} \\
e^{2\phi_1} (Q_1)^{ij} &= \sum_{k,n = 3}^d {R^i}_k {R^j}_n e^{2\phi_2} (Q_2)^{kn}, \\
{(\omega_1)_\alpha}^{ij} &= \sum_{k,n = 3}^d {R^i}_k \left({R^j}_n {(\omega_2)_\alpha}^{kn} + P^{kn} \partial_\alpha {R^j}_n\right).
\end{align}
\end{subequations}
Now, recall that by construction the coordinate vectors~$(\partial_{\lambda^i})^a$ are deviation vector fields along the foliation of extremal surfaces~$\Sigma(\lambda^i)$, and they must therefore obey the Jacobi equations~$\widetilde{J}_A (\partial_{\lambda^i})^j = 0$ for both metrics.  But from~\eqref{eqs:Gunther} it follows that any two solutions~$(\eta_A)^i$ to~$\widetilde{J}_A (\eta_A)^i = 0$ with the same boundary conditions must also be related by
\be
\label{eq:eta12}
(\eta_1)^i = \sum_{j = 3}^d {R^i}_j (\eta_2)^j,
\ee
and consequently (since the components of the~$(\partial_{\lambda^i})^a$ must be such solutions) that for all~$i,j = 3, \ldots, d$,
\be
(\partial_{\lambda^i})^a (n_1^j)_a = \sum_{k = 3}^d {R^j}_k (\partial_{\lambda^i})^a (n_2^k)_a.
\ee
Since the~$(\partial_{\lambda^i})^a$ are linearly independent, it thus follows that the bases~$\{(n_A^i)_a\}$ are related by
\be
(n_1^i)_a = \sum_{k = 3}^d {R^i}_j (n_2^j)_a.
\ee
This equation together with~\eqref{subeq:Pijconserved} implies that we may transform the basis~$\{(n_2^i)_a\}$ into the basis~$\{(n_1^i)_a\}$ without changing the components~$P^{ij}$ of the metric on the normal bundle; that is, for any~$\{(n_1^i)_a\}$, we have
\be
(g_1)^{ab} (n_1^i)_a (n_1^j)_b = (g_2)^{ab} (n_1^i)_a (n_1^j)_b.
\ee
This shows that the normal components of the inverse metric are unique:~${P^a}_b {P^c}_d (g_1)^{cd} = {P^a}_b {P^c}_d (g_2)^{cd}$ if the boundary data of~$g_1$ and~$g_2$ agree.  In particular, taking~$\{(n_1^i)_a\} = \{(d\lambda^i)_a\}$, the normal components of the inverse metrics in the coordinate system~$\{x^\alpha, \lambda^i\}$ match:~$(g_1)^{ij} = (g_2)^{ij}$.

\subsection{The $g^{\alpha i}$}
\label{subsec:off-diag}

The fact that the normal components of the metric are fixed by boundary data can be exploited to fix the off-diagonal components~$g^{\alpha i}$ as well.  To do so, consider deforming the foliation~$\Sigma(\lambda^i)$ to a one-parameter family of foliations of extremal surfaces~$\Sigma(s; \lambda^i_s)$ parametrized by~$s$, as shown in Figure~\ref{fig:shapedeformation} (we rename the parameters of the new foliation to~$\lambda^i_s$ to keep them distinct from the original parameters~$\lambda^i$, but we are imagining that for fixed~$\lambda^i$, the surface~$\Sigma(\lambda^i)$ is modified to the surface~$\Sigma(s; \lambda^i_s = \lambda_i)$).  This family of deformations is generated by a one-parameter group of diffeomorphisms~$\psi_s: M \to M$ with the property that~$\psi_s(\Sigma(\lambda^i)) = \Sigma(s; \lambda^i_s = \lambda_i)$.  These diffeomorphisms are not uniquely determined because any point on~$\Sigma(\lambda^i)$ may be mapped to any point on~$\Sigma(s; \lambda^i)$.  However, this freedom may be eliminated by recalling that both~$\Sigma(\lambda^i)$ and~$\Sigma(s; \lambda^i_s)$ admit unique isothermal coordinates~$\{x^\alpha\}$ and~$\{x^\alpha_s\}$; we may therefore fix the residual freedom in the one-parameter family of diffeomorphisms by requiring that~$\psi_s$ map the point~$p \in \Sigma(\lambda^i)$ to the point~$p_s \in \Sigma(s; \lambda^i)$ with the same isothermal coordinates as~$p$.

The group of diffeomorphisms~$\psi_s$ is generated by a vector field~$\eta^a = (\partial_s)^a$, which when restricted to each~$\Sigma(\lambda^i)$ for fixed~$\lambda^i$ can be interpreted as the deviation vector field along the one-parameter family of surfaces~$\Sigma(s; \lambda^i = \mbox{fixed})$ obtained by varying~$s$.  Moreover, since both~$\{\lambda^i, x^\alpha\}$ and~$\{\lambda^i_s, x^\alpha_s\}$ are good coordinates on~$M$, the components of~$\eta^a$ give the linearized transformation between these two coordinate systems: for each point~$p \in M$, we have
\be
\label{eq:lambdas}
\lambda^i_s(p) = \lambda^i(p) + s \eta^i(p) + \Ocal(s^2), \qquad x^\alpha_s(p) = x^\alpha(p) + s \eta^\alpha(p) + \Ocal(s^2).
\ee

\begin{figure}[t]
\centering
\includegraphics[page=7,width=0.25\textwidth]{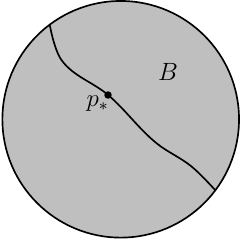}
\caption{Here we show a particular extremal surface~$\Sigma_*$ along with an arbitrarily chosen point~$p_*$ on it.  The Jacobi operator~$J$ on~$\Sigma_*$ is known from boundary data, so we may construct the linearization of the family~$\Sigma(s; \lambda^i_s)$ about~$s = 0$ by solving the Jacobi equation on~$\Sigma_*$.  Specifically, solving it only on the subdomain~$B$ whose boundary passes through~$p_*$, subject to any inhomogeneous Dirichlet boundary conditions that satisfy~$\eta^i = 0$ at~$p_*$, gives us a family~$\Sigma(s; \lambda^i_s)$ which leaves~$p_*$ fixed to first order in~$s$.}
\label{fig:Sigmastar}
\end{figure}

Next, let us consider a particular surface~$\Sigma_* \equiv \Sigma_*(\lambda^i_*)$, where~$\lambda^i_*$ is a fixed choice of parameters, and consider a point~$p_*$ on this surface; we will show that the off-diagonal metric components are unique at~$p_*$ (and therefore everywhere, since~$p_*$ and~$\Sigma_*$ were arbitrary).  To accomplish this goal, let us choose the family of foliations~$\Sigma(s; \lambda^i_s)$ such that~$p_*$ remains fixed on~$\Sigma(s; \lambda^i)$ ``to first order'' in the sense that~$\eta^i(p_*) = 0$.  A deviation vector field obeying this property can always be found by solving the Jacobi equation on a subset of~$\Sigma_*$ subject to appropriate boundary conditions.  For instance, consider the subdomain~$B \subset \Sigma_*$ shown in Figure~\ref{fig:Sigmastar}, whose boundary~$\partial B$ consists of a portion~$\partial B_\mathrm{bndry} \subset \partial \Sigma_*$ which lies on the boundary of~$\Sigma_*$ as well as a portion~$\partial B_\mathrm{bulk}$ which runs through~$\Sigma_*$ and contains the point~$p_*$.  Since the Jacobi operator~\eqref{eq:conformalJacobi} can be obtained from boundary data, we may then solve the equation~$J \eta^i = 0$ everywhere in~$B$ subject to boundary conditions that fix the~$\eta^i$ to be any specified functions on~$\partial B$, as long as~$\eta^i = 0$ at~$p_*$.  This boundary-value problem has a unique solution, and thus by construction provides us with a deviation vector~$\eta^i$ which vanishes at~$p_*$ (technically this construction provides us with a deviation vector only on~$B$, but the solution can be extended to one on all of~$\Sigma_*$ by appropriately fixing~$\eta^i$ at the remaining portion~$\partial \Sigma_* \setminus \partial B_\mathrm{bndry}$ of the boundary).  Note that there is a substantial amount of freedom in this construction: both~$B$ and~$\eta^i$ on~$\partial B$ are arbitrary, except for the condition that~$\partial B$ contain~$p_*$ and that~$\eta^i = 0$ there.

Now we may obtain the off-diagonal components of the metric.  To do so, consider the normal metric components~$g^{ij}_s \equiv g^{ab} (d\lambda^i_s)_a (d\lambda^i_s)_b$ associated to each foliation~$\Sigma(s; \lambda_s^i)$; these are computable for each~$s$ from boundary data by the procedure of Section~\ref{subsec:normal}.  Using~\eqref{eq:lambdas}, we have~$(d\lambda^i_s)_a = (d\lambda^i)_a + s \partial_a \eta^i + \Ocal(s^2)$, and thus
\bea
g^{ij}_s(x_s^\mu(p_*)) &= g^{ij}(x_s^\mu(p_*)) + 2 s g^{a(i} \partial_a \eta^{j)}|_{p_*} +  \Ocal(s^2), \\
		&= g^{ij}(x^\mu(p_*)) + s \sum_{\mu = 1}^d \left.\left[ \eta^\mu \partial_\mu g^{ij} + 2 g^{\mu(i} \partial_\mu \eta^{j)}\right] \right|_{p_*} + \Ocal(s^2).
\eea
We therefore find
\be
\label{eq:dgijds}
G^{ij} \equiv \left. \frac{d}{ds} g^{ij}_s(p_*) \right|_{s = 0} - 2 \sum_{k = 3}^d g^{k(i} \partial_k \eta^{j)}|_{p_*} = \sum_{\alpha = 1}^2 \left.\left[ \eta^\alpha \partial_\alpha g^{ij} + 2 g^{\alpha(i} \partial_\alpha \eta^{j)}\right]\right|_{p_*},
\ee
where we used the fact that by construction~$\eta^i = 0$ at~$p_*$.  This system of equations highlights the fact that the normal metric components~$g^{ij}_s$ in the perturbed foliation are given by an appropriate ``mixing'' of the normal and off-diagonal metric components~$g^{ij}$,~$g^{\alpha i}$ of the original foliation.  In particular, from Section~\ref{subsec:normal} the metric components~$g^{ij}_s$ are uniquely determined by boundary data, and therefore so is their derivative.  Since the components~$\eta^i$ are known by construction,~\eqref{eq:dgijds} relates the unknown objects~$\eta^\alpha$,~$g^{\alpha i}$ to the known quantities~$\eta^i$,~$g^{ij}$, and~$dg^{ij}_s/ds|_{s=0}$.  For this reason, we isolated all known objects into the~$G^{ij}$, and all unknowns are on the right-hand side of~\eqref{eq:dgijds}.

Equation~\eqref{eq:dgijds} is in fact a system of \textit{linear}, \textit{algebraic} equations in the unknowns~$\eta^\alpha$ and~$g^{\alpha i}$, and thus one might hope to be able to easily invert it to prove the uniqueness of the~$g^{\alpha i}$.  To assess whether or not this is possible, it is worth pausing to do some basic accounting of equations and unknowns.  In~$d$ dimensions, there are~$2(d-2)$ off-diagonal metric components~$g^{\alpha i}$ and two unknown components~$\eta^\alpha$, for a total of~$2(d-1)$ unknown quantities.  On the other hand, since there are~$(d-2)$ directions orthogonal to~$\Sigma_*$ and~$g^{ij}$ is symmetric,~\eqref{eq:dgijds} is a system of~$(d-1)(d-2)/2$ linear equations.  Therefore, for~$d \geq 6$ the number of unknowns does not exceed the number of equations, and one would expect that~\eqref{eq:dgijds} are sufficient to conclude that the~$g^{\alpha i}$ are uniquely fixed by boundary data\footnote{For~$d > 6$, the system is \textit{over}constrained, since there are more equations than unknowns, and one might be concerned that no solutions exists at all.  But recall that here our goal is to prove uniqueness: that is, we assume at least one metric satisfying~\eqref{eq:gialpha} exists, and prove that it is unique.  An overconstrained system is no obstacle to this approach. \label{foot:overconstrained}}.

On the other hand, for~$d \leq 5$ one might be concerned that the system of equations~\eqref{eq:dgijds} cannot be inverted to conclude that~$g^{\alpha i}$ are unique.  However, here we note that~\eqref{eq:dgijds} must hold \textit{for any possible choice of}~$\eta^i$ satisfying the Jacobi equation (and~$\eta^i(p_*) = 0$).  As discussed above, there is a large amount of freedom in this choice, and this additional freedom can be used to obtain more independent equations for the~$g^{\alpha i}$.  In particular, if we think of a deformed family~$\Sigma(s; \lambda^i_s)$ as a way of ``tilting'' the~$\Sigma(\lambda^i)$, then we may obtain as many ``independent'' families as there are ways of ``tilting''~$\Sigma_*$ about~$p_*$.  Since a tilt is just a rotation that mixes a direction normal to~$\Sigma_*$ and one tangent to it, there are~$2(d-2)$ independent ways of tilting~$\Sigma_*$ about~$p_*$, with each one giving rise to a version of~\eqref{eq:dgijds}.  The total number of such equations is then~$2(d-2)\times(d-1)(d-2)/2 = (d-1)(d-2)^2$, while the total number of unknowns consists of the~$2(d-2)$ components~$g^{\alpha i}$ in addition to the two unknown components~$\eta^\alpha$ of the deviation vector for each of the~$2(d-2)$ tilts, for a total of~$6(d-2)$.  We thus see that for~$d \geq 4$ there are no more equations than unknowns, and we expect that the~$g^{\alpha i}$ will be uniquely determined by known quantities.  In the case~$d = 3$, however, there are six unknowns but only two equations (from the two independent ways of tilting a two-dimensional surface in three ambient dimensions), so this straightforward method cannot be sufficient to uniquely fix the~$g^{\alpha i}$.  Indeed, the~$d = 3$ case (in Riemannian signature) was precisely the subject of~\cite{AleBal17}, which required a substantial amount of extra work to prove uniqueness of the off-diagonal terms\footnote{To prove uniqueness of the~$g^{\alpha i}$,~\cite{AleBal17} needed to use the fact that~$\{x^\alpha\}$ and~$\{x_s^\alpha\}$ are isothermal coordinates in order to derive expressions for the deviation vector components~$\eta^\alpha$.  These expressions take the form of integrals of objects containing~$g^{\alpha i}$ and~$\phi$ against Green's functions; then when inserted back into~\eqref{eq:dgijds}, one obtains two independent nonlinear \textit{integral} equations for~$g^{\alpha i}$ and~$\phi$.  Using an additional equation for~$\phi$, which is essentially the~$d = 3$ case of~\eqref{eq:extremalconstraint} below, then allows one to conclude uniqueness of the~$g^{\alpha i}$.  This approach is formidable, and requires several additional assumptions on~$M$ and~$\Sigma(\lambda^i)$ which we did not need to make here.}.  Fortunately, here we are only interested in the case~$d \geq 4$, and so we may proceed using only~\eqref{eq:dgijds}.

\begin{figure}[t]
\centering
\includegraphics[page=8,width=0.4\textwidth]{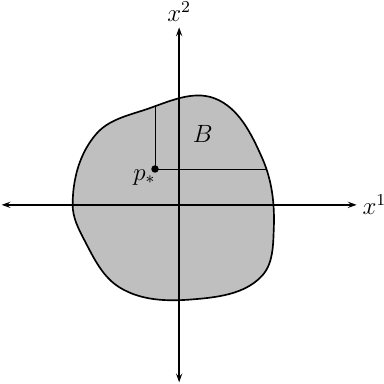}
\caption{By choosing the domain~$B$ on which we solve the Jacobi equation to be bounded by lines of fixed isothermal coordinates~$x^1$,~$x^2$ which meet at~$p_*$, we may construct~$2(d-2)$ deviation vector fields~$\eta^i$ such that for each one, precisely one of the objects~$\partial_\alpha \eta^i$ is nonzero at~$p_*$.}
\label{fig:independenttilt}
\end{figure}

To make the above intuition precise, note that the~$2(d-2)$ ``independent tilts about~$p_*$'' to which we referred are just the~$2(d-2)$ independent objects~$\partial_\alpha \eta^i$ at~$p_*$ which appear as coefficients of the~$g^{\alpha i}$ in~\eqref{eq:dgijds}.  To control these components, we construct the perturbations~$\eta^i$ as follows.  For a given point~$p_*$, take the domain~$B$ described above to be bounded within~$\Sigma_*$ by lines of constant~$x^1 = x^1(p_*)$ and~$x^2 = x^2(p_*)$, so that~$p_*$ lies on a ``corner'' of~$B$ as shown in Figure~\ref{fig:independenttilt}.  Since we may specify the components~$\eta^i$ arbitrarily on~$\partial B$ (subject to the condition~$\eta^i = 0$ at~$p_*$), we may choose all but one component of~$\eta^i$ to vanish on~$\partial B$ in a neighborhood of~$p_*$, with the remaining component (say~$\eta^d$ for simplicity) behaving like~$\eta^d|_{x^1 = x^1(p_*)} = 0$ and~$\eta^d|_{x^2 = x^2(p_*)} \propto (x^1 - x^1(p_*))$ in a neighborhood of~$B$ (the behavior of~$\eta^i$ on~$\partial B$ away from~$p_*$ is arbitrary).  Solving the Jacobi equation on~$B$ subject to these boundary conditions, we obtain a perturbation with~$\partial_1 \eta^d \neq 0$ at~$p_*$ but with all other~$\partial_\alpha \eta^i$ vanishing there.  Constructing such solutions for different choices of boundary conditions near~$p_*$, we may therefore construct~$2(d-2)$ deformations~$\eta^i_I$ (with~$I = 1, \ldots, 2(d-2)$ indexing the different perturbations) such that for each~$I$, precisely one of the objects~$\partial_\alpha \eta^i_I|_{p_*}$ is nonzero.  The equations~\eqref{eq:dgijds} then yield the~$(d-1)(d-2)^2$ equations
\be
\label{eq:GijI}
G^{ij}_I = \sum_{\alpha = 1}^2 \left.\left[ \eta_I^\alpha \partial_\alpha g^{ij} + 2 g^{\alpha(i} \partial_\alpha \eta_I^{j)}\right]\right|_{p_*} \quad \forall \, i, j = 3, \ldots, d \mbox{ and } \forall \, I = 1, \ldots, 2(d-2).
\ee
Crucially, because of the way we've constructed the perturbations~$\eta^i_I$, in each equation above (i.e.~for each~$i$,~$j$,~$I$) no more than a single off-diagonal metric component appears.  In fact,~$(d-2)$ of the equations corresponding to each~$I$ contain precisely one off-diagonal metric component, while the other~$(d-2)(d-3)/2$ contain none.  Now, write~\eqref{eq:GijI} schematically in terms of super-indices~$\mathcal{I}$,~$\mathcal{J}$ as
\be
\label{eq:schematic}
\sum_{\mathcal{J} = 1}^{6(d-2)} {R^\mathcal{I}}_\mathcal{J} V^\mathcal{J} = G^\mathcal{I} \quad \forall \, \mathcal{I} = 1, \ldots, (d-1)(d-2)^2,
\ee
where~$V^\mathcal{I}$ is a~$6(d-2)$-dimensional vector of the unknown quantities~$\eta^\alpha_I$ and~$g^{\alpha i}$,~$G^\mathcal{I}$ is a~$(d-1)(d-2)^2$-dimensional vector of the known objects~$G^{ij}_I$, and~${R^\mathcal{I}}_\mathcal{J}$ is a~$(d-1)(d-2)^2 \times 6(d-2)$ matrix of coefficients.  By construction of the~$\eta^i_I$,~$2(d-2)^2$ rows of~${R^\mathcal{I}}_\mathcal{J}$ contain precisely one nonzero entry corresponding to a coefficient of a~$g^{\alpha i}$ (though they may contain other entries corresponding to coefficients of~$\eta^\alpha_I$), while~$(d-2)^2(d-3)$ rows of~${R^\mathcal{I}}_\mathcal{J}$ contain only zeros corresponding to coefficients of~$g^{\alpha i}$.  We may then use these latter rows to row-reduce the former ones in order to eliminate any coefficients corresponding to the unknowns~$\eta_I^\alpha$.  Consequently, we obtain a row-reduced form of~${R^\mathcal{I}}_\mathcal{J}$ in which~$2(d-2)$ rows contain only precisely one nonzero entry, corresponding to each of the~$g^{\alpha i}$.  This guarantees that the~$g^{\alpha i}$ can be determined from the~$G^\mathcal{I}$, which are boundary data.  Thus we have established that the~$g^{\alpha i}$ are uniquely fixed by boundary data.

(Note that we have only ensured that the~$g^{\alpha i}$ can be recovered; we have not discussed the~$\eta^\alpha_I$ here, and indeed it is easy to see that there can be cases where the~$\eta^\alpha_I$ cannot be recovered in this way, e.g.~when~$\partial_\alpha g^{ij} = 0$ at~$p_*$.  Also note that in~$d > 4$ -- and even in~$d = 4$ in these special cases -- there will necessarily be redundancies in this procedure, since there are more equations than unknowns.  These redundancies give rise to constraints on the~$G^{ij}$ in order for the system of equations to be consistent; this is to be expected, but as remarked in footnote~\ref{foot:overconstrained}, this is not a concern for us.  Indeed, this feature is consistent with general expectations from boundary rigidity problems, which become more over-determined in higher dimensions~\cite{Stefanov:2014}.)

\subsection{The Conformal Factor}
\label{subsec:conformal}

To complete our proof, we must finally show that the induced metric on~$\Sigma$ is uniquely fixed by boundary data; in the isothermal coordinates introduced above, this amounts to showing uniqueness of the conformal factor~$\phi$.  To do so, we make use of the extremality condition~$K^a = 0$ of~$\Sigma$, but we will additionally need to assume the existence of a smooth one-parameter foliation of extremal surfaces~$\Sigma(s)$ such that~$\Sigma(s = 1) = \Sigma$ and~$\Sigma(s \to 0)$ degenerates to a point on~$\partial M$.

To proceed, consider a vector~$n^a$ normal to~$\Sigma$; the extremality condition $n_a K^a = 0$ can equivalently be expressed as~$\sigma^{ab} \grad_a n_b = 0$.  Expressing this condition in the coordinates~$\{x^\alpha, \lambda^i\}$ adapted to the foliation~$\Sigma(\lambda^i)$ containing~$\Sigma$, and moreover choosing~$n^a = (d\lambda^i)^a$ for any fixed~$i$, we find that extremality requires that the Christoffel symbols of these adapted coordinates obey
\be
0 = \sum_{\alpha,\beta = 1}^2 \sigma^{\alpha \beta} \Gamma^i_{\alpha\beta} = \sum_{\alpha,\beta = 1}^2 \sum_{\mu = 1}^d \sigma^{\alpha \beta} g^{i\mu} \left[\partial_\alpha g_{\mu\beta} - \frac{1}{2} \partial_\mu g_{\alpha\beta} \right] \quad \forall i = 3, \ldots, d,
\ee
where the second equality is obtained by just using the usual formula for the Christoffel symbols.  Now, the components of the induced metric in these isothermal coordinates are just~$\sigma_{\alpha\beta} = g_{\alpha\beta} = e^{2\phi} \delta_{\alpha\beta}$, and thus~$\sigma^{\alpha\beta} = e^{-2\phi} \delta^{\alpha\beta}$ (note that in general,~$\sigma^{\alpha\beta} \neq g^{\alpha\beta}$).  Using these expressions and decomposing the full sum over~$\mu$ into partial sums in the normal and tangent directions to~$\Sigma$, the above constraint simplifies to
\be
\sum_{j = 3}^d g^{ij} \left[\sum_{\alpha,\beta = 1}^2 \delta^{\alpha\beta} \partial_\alpha g_{j\beta} - 2 e^{2\phi} \partial_j \phi \right] = 0 \quad \forall i = 3, \ldots, d.
\ee
Next, note that from~\eqref{eqs:ginverse}, we concluded that the~$(d-2) \times (d-2)$ matrix with entries~$g^{ij}$ is invertible; this means that the free index~$i$ above can be lowered by multiplying by this inverse, and we obtain
\be
\sum_{\alpha,\beta = 1}^2 \delta^{\alpha\beta} \partial_\alpha g_{i\beta} - 2 e^{2\phi} \partial_i \phi = 0 \quad \forall i = 3, \ldots, d.
\ee
Finally, using~\eqref{eq:gialpha} to express the metric components~$g_{\beta j}$ in terms of~$\phi$ and the unique components~$g^{\alpha i}$,~$g^{ij}$, we finally obtain
\be
\label{eq:extremalconstraint}
\sum_{\alpha = 1}^2 \left(\partial_\alpha f_{\alpha i} + 2 f_{\alpha i} \partial_\alpha \phi\right) - 2\partial_i \phi = 0 \quad \forall i = 3, \ldots, d,
\ee
where we recall that~$f_{\alpha i}$ is a known function of the inverse metric components~$g^{\alpha i}$ and~$g^{ij}$.  Since we have already established that these components are uniquely fixed by boundary data,~$f_{\alpha i}$ is as well.

Equation~\eqref{eq:extremalconstraint} is sufficient to prove uniqueness of~$\phi$.  To see this, recall that we have assumed that one of the parameters~$\lambda^i$, say~$\lambda^d$, gives rise to a continuous foliation of extremal surfaces that limits to an arbitrarily small surface near~$\partial M$ as~$\lambda^d \to 0$.  Now, fix all the~$\lambda^{i \neq d} = 0$ and let~$\Xi$ be the surface foliated by~$\Sigma(\lambda^{i \neq d} = 0, \lambda^d)$ as~$\lambda^d$ is varied; then equation~\eqref{eq:extremalconstraint} with~$i = d$ is an inhomogeneous linear first-order PDE for~$\phi$ on~$\Xi$; since the~$f_{\alpha i}$ are uniquely fixed by boundary data, this equation is unique.  This equation is hyperbolic, which means in particular that as long as boundary conditions on~$\partial M$ are provided, it can be evolved forward from~$\lambda^d = 0$ to all of~$\Xi$.  But since the boundary value of~$\phi$ is fixed boundary data,~$\phi$ is fixed by boundary data everywhere on~$\Xi$ as well.  Thus~$\phi$ is unique.

\section{Discussion}
\label{sec:conc}

As a standalone geometric result, the argument presented in this paper extends the work of~\cite{AleBal17} and related boundary rigidity literature to higher (co)dimension and general signature bulk geometries.  Our primary interest, however, is in its application in the context of AdS/CFT, in which our result implies that the metric of an AlAdS bulk is uniquely fixed by second variations of the areas of a smooth family of two-dimensional boundary-anchored extremal surfaces.  We emphasize again that, in contrast with many of the approaches to bulk metric reconstruction in the literature, our result does not assume any symmetries of the bulk geometry, and in particular applies to generic dynamical geometries.  In the special case of AdS$_4$/CFT$_3$, such boundary-anchored two-dimensional extremal surfaces arise naturally from the HRT conjecture, and their areas are interpreted as entanglement entropies\footnote{Only when the extremal surfaces in question are the minimal-area ones homologous to the boundary regions to which they're anchored, though see e.g.~\cite{EngWal17,EngWal18} for holographic interpretations of non-minimal-area extremal surfaces.}.  We now comment on some limitations, generalizations, and potential future directions.

\subsubsection*{Towards an Explicit Reconstruction}

The argument we have presented here is a uniqueness result: we have shown that the bulk metric (if one exists) is uniquely fixed by boundary data.  Of course, it would be much more desirable to have a constructive result in the form of an algorithmic way of obtaining the bulk metric from the boundary data (which, moreover, should be tractable).  What would be needed to obtain such an explicit reconstruction?

Of the four steps (outlined in Section~\ref{subsec:sketch}) in our argument, two are not constructive.  First, the argument that a consistent set of isothermal coordinates~$\{x^\alpha\}$ exists on each~$\Sigma$ is obtained by a proof by contradiction, and thus does not actually construct the coordinate system~$\{x^\alpha\}$ explicitly.  Second, in order to show that the normal metric components~$g^{ij}$ are uniquely fixed by boundary data, we invoked the uniqueness theorem~\cite{AlbGui13} to conclude that the Jacobi operator~$J$ on each~$\Sigma$ is fixed (up to gauge) by second area variations.  On the other hand, if the coordinate system~$\{x^\alpha\}$ and the~$g^{ij}$ are somehow obtained explicitly, the final two steps in our argument -- obtaining the off-diagonal metric components~$g^{i\alpha}$ and the conformal factor~$\phi$ -- are constructive in that they require only solving a system of linear algebraic equations or a first-order hyperbolic PDE with known coefficients.

Extending our result to an explicit reconstruction therefore requires a constructive way way of obtaining the~$\{x^\alpha\}$ and of recovering the Jacobi operator from its Cauchy data.  We do not give suggestions for how to accomplish the former, but extant results in the literature suggest that a solution to the latter should be possible to obtain.  In particular, consider the operator~$\Delta + Q$ acting on some domain of~$\mathbb{R}^2$, where~$\Delta$ is the usual flat-space Laplacian; it is shown in~\cite{NovikovSantacesaria} that the potential~$Q$ can be \textit{explicitly} reconstructed from the Cauchy data of~$\Delta + Q$.  Now, in our context recall that the Jacobi operator can be written as~$\widehat{D}^\dag \widehat{D} + Q$, where~$\widehat{D}$ is a nontrivial connection and~$\widehat{D}^\dag$ is its adjoint with respect to the inner product~\eqref{eq:inprod}.  Because the connection~$\widehat{D}$ is nontrivial, the result of~\cite{NovikovSantacesaria} unfortunately does not allow us to recover the Jacobi operator from boundary data.  However, the similarity between these two problems, and the fact that an explicit reconstruction formula does exist for the simpler operator~$\Delta + Q$, suggests to us that it should be possible to obtain an analogous reconstruction formula for operators of the form~$\widehat{D}^\dag \widehat{D} + Q$ as well.  Presumably such a formula would require some kind of gauge-fixing of the basis~$\{(n^i)_a\}$, which could be enforced, for instance, by the condition
\be
n^i\cdot(\partial_{\lambda^j})= {\delta^i}_j,
\ee
which fixes~$(n^i)_a = (d\lambda^i)_a$.  An explicit reconstruction formula for the connection~$\widehat{D}$ (i.e.~for the connection coefficients~${\omega_{\alpha i}}^j$) in this gauge would then allow us to explicitly recover the components~$g^{ij}$ of the metric via the compatibility condition
\be
\label{eq:pfaff}
\widehat{D}_\alpha g^{ij} = \partial_\alpha g^{ij}-2\sum_{k=3}^d {\omega_{\alpha k}}^{(i} g^{j)k} = 0;
\ee
by Theorem~4.2 of~\cite{Mar07} this equation has at most one solution with given boundary conditions, which can be obtained by integrating~\eqref{eq:pfaff} in from the boundary along any path.

It therefore seems that obtaining a constructive version of the results in this paper is not beyond reach; we leave this as a promising direction of future work\footnote{Alternatively, it may be possible to use the tensor Radon transform to reconstruct bulk metrics from boundary data. This line of thinking seems more commonplace in tomographic applications where numerical efficiency is paramount.  However, successful reconstruction using these approaches seems to rely on having chosen a reasonable background metric \textit{a priori}.}.

\subsubsection*{A Diagnostic on the Existence of a Dual Geometry}

Our result relies on the uniqueness to several substantially overconstrained systems of equations (for instance, the algebraic system of equations~\eqref{eq:schematic}, or the hyperbolic PDE~\eqref{eq:extremalconstraint}, which can be defined on any one-parameter subfamily of the foliation~$\Sigma(\lambda^i)$).  This overdetermination is a general expectation in inverse boundary value problems: after all, the space of all possible subregions of the boundary (and hence the space of all possible boundary-anchored minimal surfaces) is much larger than the space of bulk metric components.  In fact, if a constructive version of our argument can be developed, then this overdetermination suggests a way of diagnosing when a CFT state does not admit a bulk dual geometry: if any of these overdetermined systems of equations do not admit a solution, then a bulk geometry cannot exist.  Indeed, a similar observation was made in~\cite{EngFis17,EngFis17b} in the context of bulk reconstruction from light-cone cuts.  More generally, this question could be nontrivially related to other known constraints on what classes of CFT states are permitted to have classical bulk gravity duals, such as the entropic constraints of~\cite{bao2015holographic, hayden2013holographic}.

\subsubsection*{Maximum Bulk Depth}

Our result relies crucially on the existence of the foliation of extremal surfaces~$\Sigma(\lambda^i)$ with disk topology, which in turn constrains the region~$\Rcal$ of the bulk to which our result applies.  For instance, in a static black hole geometry, extremal surfaces anchored to a connected boundary region never enter the event horizon, and thus any such foliation~$\Sigma(\lambda^i)$ cannot probe inside the black hole.  If our goal is to probe the geometry in regions of strong gravity, how concerning is this observation?  This question is especially relevant in light of the no-go theorem of~\cite{Engelhardt:2015dta}, which forbids any metric reconstruction approach based on ``hole-ography'' from recovering the metric in any region of sufficiently strong gravity.

While it is true that a typical topologically nontrivial spacetime cannot be foliated completely by a family~$\Sigma(\lambda^i)$ of the type we require, it is nevertheless the case that there exist generic examples of spacetimes in which a family~$\Sigma(\lambda^i)$ can penetrate beyond the event horizon, and even past apparent horizons.  For example, the extremal surfaces considered in~\cite{Liu:2013iza,Liu:2013qca,Hubeny:2013dea} in Vaidya-AdS are able to penetrate both the early-time event and apparent horizons of a black hole formed from collapse, and thus we conclude that the metric in those regions is indeed uniquely fixed by boundary data (this result evades the no-go theorem of~\cite{Engelhardt:2015dta} because ``hole-ographic'' reconstructions require the existence of extremal surfaces that are tangent to arbitrarily small closed spacelike curves that degenerate to a point).

Relatedly, subregion/subregion duality also makes it natural to ask how much of the metric in the entanglement wedge~$W_E[R]$ of a boundary subregion~$R$ is ensured to be unique by our argument.  In other words: how much of~$W_E[R]$ can be foliated by a continuous family~$\Sigma(\lambda^i)$ of extremal surfaces which are anchored only to~$R$?  Recall that the boundary~$\partial W_E[R]$ is generated by null congruences fired from the HRT surface~$X[R]$, which in a generic spacetime encounter caustics and self-intersect, leading to ``cusps'' in~$\partial W_E[R]$.  Because extremal surfaces are smooth, and because the family~$\Sigma(\lambda^i)$ must be continuous, generically we would expect that~$W_E[R]$ should always contain regions near these cusps that cannot be accessed by any family~$\Sigma(\lambda^i)$ living only in~$W_E[R]$.

\subsubsection*{Going Deeper?}

An interesting future direction would be to consider whether this proof method can be used to reconstruct previous ``shadow'' regions in the bulk via surface-state correspondence \cite{miyaji2015surface} and entanglement of purification \cite{takayanagi2017holographic, nguyen2018entanglement} techniques. The entanglement of purification conjecture allows us access to areas of bulk, rather than boundary, anchored minimal surfaces, while the surface-state correspondence helps to recast these bulk-anchored minimal surfaces as boundary-anchored minimal surfaces of a transformed boundary. Combined with our results, these bulk-anchored surfaces could be used in an iterative fashion to build the metric further into the spacetime than boundary-anchored surfaces can reach, in a method complementary to the tensor network-based techniques of \cite{BaoPen18}.  Even boundary-anchored surfaces can already reach past event horizons, as in the case of Vaidya spacetimes, so it is natural to ask just how far these bulk-anchored surfaces reach. Furthermore, it would be interesting to ask how multipartite and conditional entanglements of purification \cite{umemoto2018entanglement, bao2018holographic, bao2018conditional} can also help to probe deeper into the bulk.

\subsubsection*{Properties of~$\partial M$}

Our result is purely geometric, and applies to any geometry with finite boundary~$\partial M$: in other words, whenever the aforementioned boundary data is known (including most importantly the areas of two-dimensional spacelike extremal surfaces anchored to~$\partial M$), so is the metric in the region~$\Rcal$.  Indeed, it has been suggested (though we remain agnostic on this topic) that this boundary data might be available for holographic screens in the program of generalized holography~\cite{Sanches:2016sxy,Nomura:2016ikr}; if true, our result would apply.

In a related direction, because we took the boundary~$\partial M$ to be a finite boundary, we interpreted our results as applying to a \textit{regulated} AlAdS geometry.  It would perhaps be more elegant, however, to reformulate our results in terms of renormalized (rather than merely regulated) quantities evaluated at the \textit{bona fide} asymptotic boundary.  At least in~$d = 4$ it is clear how this might be done: as shown in~\cite{FisWis16}, the renormalized area of a boundary-anchored two-dimensional extremal surface~$\Sigma$ of disk topology can be computed by the functional
\be
A_\mathrm{ren}[\Sigma] = \int_\Sigma \left(1 + \frac{\ell^2}{2} \, ^\Sigma \! R\right) - 2\pi \ell^2,
\ee
where~$\ell$ is the AdS length and~$^\Sigma \! R$ is the Ricci scalar of~$\Sigma$.  Since the additional term in the integral is just topological (by the Gauss-Bonnet theorem), we expect that variations of~$A_\mathrm{ren}$ (and therefore also of the renormalized entangelement entropy) should still yield the Cauchy data~$\Ccal_J$, from which our result would still follow.  It would be worth verifying in more detail that this expectation is indeed correct, as well as checking whether an analogous construction exists in more general dimension.

\subsubsection*{More General Spacetime Emergence}

Another potentially interesting connection is with the emergent spacetime program of~\cite{Cao:2016mst,Cao:2017hrv}. For instance,~\cite{Cao:2017hrv, Monard2015} provides a way to reconstruct the metric tensor of the near-flat emergent geometry using inverse tensor Radon transforms. It is natural to ask if our methods can be used to derive the metric tensor from area data but without prior knowledge of or restrictions on the background, and if this inversion can be done in a covariant way. 

\subsubsection*{Higher Dimensions}

Here we focused exclusively on the case of two-dimensional extremal surfaces.  Restricting to two dimensions allowed us to invoke the isothermal coordinates~$\{x^\alpha\}$, and also allowed our use of the uniqueness result~\cite{AlbGui13} to fix the Jacobi operator from boundary data.  Of course, it is natural to ask whether the general proof technique can be extended beyond two-dimensional surfaces.  The answer is unclear; besides a generalization of~\cite{AlbGui13}, we suspect the most difficult task would be to fix a set of coordinates on the~$\Sigma(\lambda^i)$ that put the induced metric~$\sigma_{ab}$ in some fiducial form analogous to the isothermal form~\eqref{eq:isothermal}.  We leave this investigation to future work.

\subsubsection*{Quantum Corrections}

We have so far exclusively discussed the question of recovering a \textit{classical} bulk geometry from boundary data.  But the importance of this task in AdS/CFT stems from the fact that such a recovery should eventually lead to a way of probing quantum corrections to the bulk gravitational theory.  It is worth asking, therefore, whether the approach we have followed here is amenable to a quantum generalization.  To that end, first recall that under quantum corrections, the HRT prescription is modified: the entanglement entropy of a boundary region~$R$ is no longer given by the area of the bulk extremal surface~$X[R]$ homologous to~$R$, but rather by the generalized entropy of the bulk \textit{quantum} extremal surface~$\Xcal[R]$ homologous to~$R$~\cite{EngWal14},
\be
S[R] = S_\mathrm{gen}[\Xcal[R]] \equiv \frac{\Area[\Xcal[R]]}{4G_N \hbar} + S_\mathrm{out}[\Xcal[R]],
\ee
where~$\Xcal[R]$ is a stationary point of the functional~$S_\mathrm{gen}[\Xcal]$ and~$S_\mathrm{out}[\Xcal]$ is the (bulk) entropy of any quantum fields outside of~$\Xcal$.  The challenge in extending our results to include the quantum correction term~$S_\mathrm{out}[\Xcal]$ is thus to understand whether switching from classical extremal surfaces to quantum extremal surfaces spoils the uniqueness argument.  In particular, we must generalize the Jacobi equation~\eqref{eq:formalJacobi}, which governs the behavior of the family of classical extremal surfaces, to a quantum version governing the behavior of a family of \textit{quantum} extremal surfaces.  Fortunately, this generalization is provided in~\cite{EngFis19}, where the quantum correction to~\eqref{eq:formalJacobi} is shown to take the form of functional derivatives of~$S_\mathrm{out}$.  While a ``quantum generalization'' of the uniqueness theorem of~\cite{AlbGui13} is presumably not currently known, we expect that the precise quantum-corrected formalism of~\cite{EngFis19} should provide a way of systematically tracking through these quantum corrections.  We leave this as a direction of future work.

\section*{Acknowledgements}

We thank Spyridon Alexakis, Xi Dong, Netta Engelhardt, Nikolaos Eptaminitakis, Daniel Harlow, Gary Horowitz, Robin Graham, Daniel Kabat, Sorin Mardare, Reed Meyerson, and Matteo Santacesaria for useful and interesting conversations while this work was being completed; we are especially grateful to Tracey Balehowsky for her patience in explaining her work~\cite{AleBal17} to us and to Gunther Uhlmann for extensive discussions regarding inverse boundary value problems.  CC would also like to thank Gunther Uhlmann for his helpful suggestions and his hospitality during the visits to the University of Washington.  This research was supported in part by the National Science Foundation under Grant No.~NSF PHY-1748958; in particular, NB, SF, and CK would like to thank both the KITP and OIST for hospitality during the completion of a portion of this work.
NB is supported by the National Science Foundation under grant number 82248-13067-44-PHPXH, by the Department of Energy under grant number DE-SC0019380, and by New York State Urban Development Corporation – Empire State Development – contract no. AA289. CC acknowledges the support by the U.S. Department of Energy, Office of Science, Office of High Energy Physics, under Award Number DE-SC0011632, as well as by the U.S. Department of Defense and NIST through the Hartree Postdoctoral Fellowship at QuICS. SF acknowledges the support of the Natural Sciences and Engineering Research Council of Canada (NSERC), funding reference number SAPIN/00032-2015.  The work of CK is supported by the U.S.~Department of Energy under grant number DE-SC0019470.  This work was supported in part by a grant from the Simons Foundation (385602, AM).

\appendix

\section{Area Variation Formulas}
\label{app:areavariation}

Here we compile the formulas used in the main text for first and second variations of the areas of extremal surfaces.  First, let us note that for any one-parameter family of spacelike surfaces~$\Sigma(s)$, any integrated object
\be
F(s) = \int_{\Sigma(s)} f \, \bm{\eps}
\ee
(where~$\bm{\eps}$ is the natural volume form on~$\Sigma(s)$) has derivative (see e.g.~Appendix C of~\cite{FisWis16})
\be
\label{eq:Fprime}
F'(0) = \int_{\Sigma(0)} \left(\eta^a_\perp \grad_a f + \eta_a K^a f \right) \, \bm{\eps} + \int_{\partial \Sigma(0)} N_a \eta^a f \, ^\partial \! \bm{\eps},
\ee
where~$\eta^a = (\partial_s)^a|_{s = 0}$ is deviation vector along this family of surfaces,~$\eta^a_\perp$ is its component normal to~$\Sigma(0)$,~$K^a$ is the mean curvature of~$\Sigma(0)$,~$N^a$ is the unit outward-pointing normal to~$\partial \Sigma(0)$ in~$\Sigma(0)$, and~$\, ^\partial \! \bm{\eps}$ is the natural volume form on~$\partial \Sigma(0)$.

Now take the~$\Sigma(s)$ to all be extremal codimension-two surfaces, so that~$K^a = 0$, and also take~$f = 1$, so that~$F(s)$ just computes the area of~$\Sigma(s)$.  Then it is clear that a first derivative of the area is sensitive only to a boundary term,
\be
A'(0) = \int_{\partial \Sigma(0)} N_a \eta^a \, ^\partial \! \bm{\eps},
\ee
reproducing the first area variation formula~\eqref{eq:firstareavariation} in the main text.

To obtain a second area variation formula, consider a two-parameter family of extremal surfaces~$\Sigma(s_1,s_2)$ and apply~\eqref{eq:Fprime} twice:
\bea
\left.\frac{\partial^2 A(s_1,s_2)}{\partial s_2 \, \partial s_1}\right|_{(s_1,s_2) = (0,0)} &= \left.\frac{\partial}{\partial s_2} \left(\int_{\partial \Sigma(s_1,s_2)} N_a \eta_1^a \, ^\partial \! \bm{\eps} \right)\right|_{(s_1,s_2) = (0,0)}, \\
		&= \int_{\partial \Sigma(0,0)} \left(\eta_2^b \grad_b (N_a \eta_1^a) + (\eta_2)_b k^b N_a \eta_1^a\right) \, ^\partial \! \bm{\eps}, \label{eq:Apprime}
\eea
where~$k^a$ is the mean curvature of~$\partial \Sigma(0,0)$ in~$\partial M$ and without loss of generality we are taking~$\eta_1^a$ and~$\eta_2^a$ to be normal to~$\partial \Sigma(0,0)$.  Now we write the first term as
\be
\label{eq:temp1}
\eta_2^a \grad_a (N_b \eta_1^b) = N^a \left[\eta^b_2 \grad_b (\eta_1)_a + (\eta_1)_b \grad_a \eta_2^b\right] + (\eta_1)_a \pounds_{\eta_2} N^a
\ee
and simplify it piece-by-piece.  First, note that we can decompose~$\eta_1^a = \eta^a_{1,\perp} + \eta^a_{1,\parallel}$, where~$\eta^a_{1,\parallel} = {\sigma_b}^a \eta_1^b$ is the component of~$\eta_1^a$ tangent to~$\Sigma(0,0)$; since~$\eta_1^a$ is normal to~$\partial \Sigma(0,0)$, we must have~$\eta^a_{1,\parallel} = (N \cdot \eta_1) N^a$ at~$\partial \Sigma(0,0)$.  Using the same expansion for~$\eta_2^a$, we have
\bea
N^a (\eta_1)_b \grad_a \eta_2^b &= \left[(\eta_{1,\perp})_b + (N \cdot \eta_1) N_b \right] N^a \grad_a \left[\eta_{2,\perp}^b + \eta_{2,\parallel}^b\right], \\
		&= (\eta_{1,\perp})_b N^a \grad_a \eta_{2,\perp}^b + (\eta_{1,\perp})_b N^a \grad_a (\eta^b_{2,\parallel}) + (N \cdot \eta_1) N_b N^a \grad_a \eta_2^b, \\
		&= \sum_{i = 1}^{d-2} \eta_1^i N^a \widehat{D}_a (\eta_2)_i - (N \cdot \eta_2) N_b N^a \grad_a \eta_{1,\perp}^b + (N \cdot \eta_1) N_b N^a \grad_a \eta_2^b, \label{eq:temp3}
\eea
where to get to the last line we decomposed~$\eta_{1,\perp}^a$ in the basis~$\{(n^i)^a\}$ of the normal bundle.  The middle term can be simplified as
\be
\label{eq:temp2}
N_b N^a \grad_a \eta_{1,\perp}^b = N^a N^b {K^c}_{ab} (\eta_1)_c = -k^a (\eta_{1,\perp})_a,
\ee
where we used the definition of the extrinsic curvature tensor~${K^c}_{ab}$ and the fact that
\be
N^a N^b {K^c}_{ab} (\eta_1)_c = (\sigma^{ab} - y^{ab}) {K^c}_{ab} (\eta_1)_c = (K^c - k^c)(\eta_{1,\perp})_c = -k^c (\eta_{1,\perp})_c,
\ee
where~$y^{ab}$ is the induced metric on~$\partial \Sigma(0,0)$ and~$K^c = 0$ by extremality.

Next, to evaluate the Lie derivative term~$(\eta_1)_a \pounds_{\eta_2} N^a$, we first note that the fact that~$N^a$ is a unit vector implies that
\be
\pounds_{\eta_2} (N_a N_b g^{ab}) = 0 \Rightarrow N^a \pounds_{\eta_2} N_a = N^a N^b \grad_a (\eta_2)_b.
\ee
The fact that~$N^a$ is tangent to~$\Sigma(s_1,s_2)$ implies that its Lie derivative must be as well:~$\pounds_{\eta_2} N^a = {\sigma^a}_b \pounds_{\eta_2} N^b$.  Finally, the fact that~$N^a$ is normal to~$\partial \Sigma(s_1,s_2)$ implies that~$N^a = p^{ab}N_b$, where~$p^{ab} = g^{ab} - y^{ab}$ is the normal projector to~$\partial \Sigma(s_1,s_2)$.  Using these properties, along with the fact that the Lie derivative along the induced metrics is\footnote{The Lie derivative~$\pounds_{\eta_2} \sigma^{ab}$ can be computed relatively easily by realizing that the tangent space to~$\Sigma$ is metric-independent, so~${\sigma_a}^b$ must act as the identity on any vector tangent to~$\Sigma$, and thus the Lie derivative~$\pounds_{\eta_2} {\sigma_a}^b$ must vanish when its lower index is projected onto~$\Sigma$.  Then using~$\pounds_{\eta_2} \sigma^{ab} = \pounds_{\eta_2}({\sigma^a}_c {\sigma^b}_d g^{cd})$, one obtains the above expression.  More details on this derivation will be provided in~\cite{EngFis19}.}
\be
\pounds_{\eta_2} \sigma^{ab} = -2\sigma^{ac} \sigma^{bd} \grad_{(c} (\eta_2)_{d)}, \qquad 
\pounds_{\eta_2} y^{ab} = -2y^{ac} y^{bd} \grad_{(c} (\eta_2)_{d)},
\ee
it is straightforward to show that
\be
\pounds_{\eta_2} N^a = {\sigma^a}_b \pounds_{\eta_2} N^b = {\sigma^a}_b \pounds_{\eta_2} \left(p^{bc} N_c\right) = -(\sigma_{ab} + y_{ab}) N_c \grad^{(b} \eta_2^{c)}.
\ee
We thus obtain
\be
\label{eq:temp4}
(\eta_1)_a \pounds_{\eta_2} N^a = -(N \cdot \eta_1) N^a N_b \grad_a \eta_2^b.
\ee

Inserting~\eqref{eq:temp2} into~\eqref{eq:temp3},~\eqref{eq:temp3} and~\eqref{eq:temp4} into~\eqref{eq:temp1}, and finally~\eqref{eq:temp1} into~\eqref{eq:Apprime}, we finally obtain (after minor cancellations and rearrangement)
\begin{multline}
\left.\frac{\partial^2 A(s_1,s_2)}{\partial s_2 \, \partial s_1}\right|_{(s_1,s_2) = (0,0)} = \int_{\partial \Sigma(0,0)} \left[\sum_{i = 1}^{d-2} \eta_1^i N^a \widehat{D}_a (\eta_2)_i \right. \\ \left. \phantom{\sum_{i = 1}^s} + N_a \eta_2^b \grad_b \eta_1^a + 2(N \cdot \eta_{(1}) (\eta_{2)})_a k^a - (N \cdot \eta_1) (N \cdot \eta_2) k^a N_a \right] \, ^\partial \! \bm{\eps},
\end{multline}
which is equation~\eqref{eq:secondareavariation} quoted in the text.

\section{Boundary Values of $\eta^i$}
\label{app:boundaryeta}

Consider the foliation of extremal surfaces~$\Sigma(\lambda^i)$; the existence of this foliation means that the parameters~$\lambda^i$ are scalars both in~$M$ and~$\partial M$; we may then introduce a basis of the normal bundle of~$\Sigma$ in~$M$ as~$(n^i)^a = (d\lambda^i)^a$, and a basis of the normal bundle of~$\partial \Sigma$ in~$\partial M$ as~$(\tilde{n}^i)^a = h^{ab} (d\lambda^i)_b$, where the tilde denotes a boundary object.  The basis~$\{(\tilde{n}^i)^a\}$ is known on the boundary, and therefore for any deviation vector~$\eta^a$ on~$\Sigma$ such that~$\eta^a|_{\partial \Sigma}$ is tangent to~$\partial M$, the components~$\tilde{\eta}^i \equiv \tilde{n}^i \cdot \eta|_{\partial \Sigma}$ are as well.  However, since~$\eta^a|_{\partial \Sigma}$ is tangent to the boundary, we have that
\be
\eta^i|_{\partial \Sigma} = \eta^a (d\lambda^i)_a|_{\partial \Sigma} =  \eta^a {h_a}^b (d\lambda^i)_b|_{\partial \Sigma} = \tilde{\eta}^i,
\ee
and thus with this particular choice of basis, the components~$\eta^i|_{\partial \Sigma}$ are in fact known boundary data as well.  We may also show that~$g^{ij}|_{\partial \Sigma}$ are known in this basis as follows.  Writing the induced metric on~$\partial M$ as
\be
h_{ab} = g_{ab}|_{\partial M} - v_a v_b
\ee
(where as in the main text~$v^a$ is the unit normal to~$\partial M$ in~$M$), we may contract this equation with~$(n^i)^a$ and~$(n^j)^b$ to obtain
\be
\label{eq:tildegij}
\tilde{g}^{ij} = g^{ij}|_{\partial \Sigma} - (v \cdot n^i) (v \cdot n^j),
\ee
where~$\tilde{g}^{ij} \equiv \tilde{n}^i \cdot \tilde{n}^j = h_{ab} (n^i)^a (n^j)^b$.  Now, since~$v^a$ is normal to~$\partial M$, it is also normal to~$\partial \Sigma$, and we may therefore decompose it in the basis~$\{N^a, (\tilde{n}^i)^a\}$ of the normal bundle of~$\partial \Sigma$ in~$M$.  It is straightforward to show, using the fact that~$v^a$ is a unit vector, that this decomposition takes the form
\be
v^a = \frac{1}{\sqrt{1-h_{ab} N^a N^b}} \left[N^a - \sum_{i,j = 1}^{d-2} \tilde{g}_{ij} \widetilde{N}^i (\tilde{n}^j)^a\right],
\ee
where~$\widetilde{N}^i = h_{ab} N^a (\tilde{n}^i)^b$ is known from first area variations, since such variations determine the projection~${h^a}_b N^b$.  Using this decomposition to evaluate~$v \cdot n^i$ in~\eqref{eq:tildegij}, we ultimately find
\be
g^{ij}|_{\partial M} = \tilde{g}^{ij} + \frac{\widetilde{N}^i \widetilde{N}^j}{1-h_{ab} N^a N^b},
\ee
which gives the promised expression for the components~$g^{ij}|_{\partial M}$ in terms of boundary data.  Inverting~$g^{ij}|_{\partial \Sigma}$ gives~$g_{ij}|_{\partial \Sigma}$, and it then follows in particular that the components~$\eta_i|_{\partial \Sigma} = \sum_j g_{ij} \eta^j|_{\partial \Sigma}$ can also be expressed in terms of boundary data:
\be
\eta_i|_{\partial \Sigma} = \tilde{\eta}_i - (N \cdot \eta) \widetilde{N}_i,
\ee
where normal bundle indices on tilded objects are raised and lowered with~$\tilde{g}_{ij}$ and~$\tilde{g}^{ij}$.

\bibliographystyle{jhep}
\bibliography{all}

\end{document}